%% file: cQED_DW_chem_JPCL_final.tex
\documentclass[journal=jpclcd,manuscript=letter]{achemso}
\mciteErrorOnUnknownfalse
\usepackage{amsmath, amsfonts}
\usepackage{booktabs}
\usepackage[dvipsnames]{xcolor}
\usepackage{hyperref}

\usepackage{soul}

\setkeys{acs}{maxauthors = 0} % necessary to have no truncation on bibliographical entries when using the achemso

\def\hH{ \hat{H} }
\def\hkc{ \hat{H}_{\mathrm{KC}} }
\def\hdw{ \hat{H}_{\mathrm{DW}} }
\def\ha{ \hat{a} }
\def\hadag{ \ha^{\dagger} }

\def\hx{ \hat{x} }
\def\hp{ \hat{p} }
\def\hr{ \hat{\rho} }
\def\dhr{ \dfrac{\partial \hr}{\partial t} }

\def\nth{ n _{\text{th}}}
\def\lind{\hat{\mathcal{L}}}
\def\nnn{ \nonumber \\ }
\newcommand{\lrp}[1]{ \left( {#1} \right) }
\newcommand{\lrb}[1]{ \left[ {#1} \right] }
\newcommand{\lrB}[1]{ \left\{ {#1} \right\} }

\author{Delmar G. A. Cabral}
\affiliation{Department of Chemistry, Yale 
University, New Haven, CT 06520, USA}
\altaffiliation{Contributed equally to this work} 
\author{Pouya Khazaei} \affiliation{Department of Chemistry, University of 
Michigan, Ann Arbor, MI 48109, USA}
\altaffiliation{Contributed equally to this work} 
\author{Brandon C. Allen}
\affiliation{Department of Chemistry, Yale 
University, New Haven, CT 06520, USA}
\altaffiliation{Contributed equally to this work} 
\author{Pablo E. Videla}
\affiliation{Department of Chemistry, Yale University, New Haven, CT 06520,
USA}
\altaffiliation{Contributed equally to this work} 
\author{Max Sch\"afer} \affiliation{Department of Applied Physics and 
Physics, Yale University, New Haven, CT 06520, USA}
\alsoaffiliation{Yale Quantum Institute, Yale University, New Haven, CT 06511,
USA}
\author{Rodrigo G. Corti\~{n}as}\affiliation{Department of Applied Physics and 
Physics, Yale University, New Haven, CT 06520, USA}
\alsoaffiliation{Yale Quantum Institute, Yale University, New Haven, CT 06511,
USA}
\author{Alejandro Cros Carrillo de Albornoz}\affiliation{Department of Applied
Physics and 
Physics, Yale University, New Haven, CT 06520, USA}
\alsoaffiliation{Yale Quantum Institute, Yale University, New Haven, CT 06511,
USA}
\alsoaffiliation{Department of Physics and Astronomy, University College
London, London WC1E 6BT, UK}
\author{Jorge Ch\'{a}vez-Carlos} \affiliation{Department of Physics, University 
of Connecticut, Storrs, CT 06511, USA}
\author{Lea F. Santos} \affiliation{Department of Physics, University of 
Connecticut, Storrs, CT 06511, USA}
\author{Eitan Geva} \affiliation{Department of Chemistry, University of 
Michigan, Ann Arbor, MI 48109, USA}
\email{eitan@umich.edu}
\author{Victor S. Batista}
\affiliation{Department of Chemistry, Yale University, New Haven, CT 06520,
USA}
\alsoaffiliation{Yale Quantum Institute, Yale University, New Haven, CT 06511, 
USA}
\email{victor.batista@yale.edu}

\title{A Roadmap for Simulating Chemical Dynamics on a Parametrically Driven Bosonic Quantum Device}

\begin{document}

\begin{tocentry}
    \includegraphics[width=1.0\textwidth]{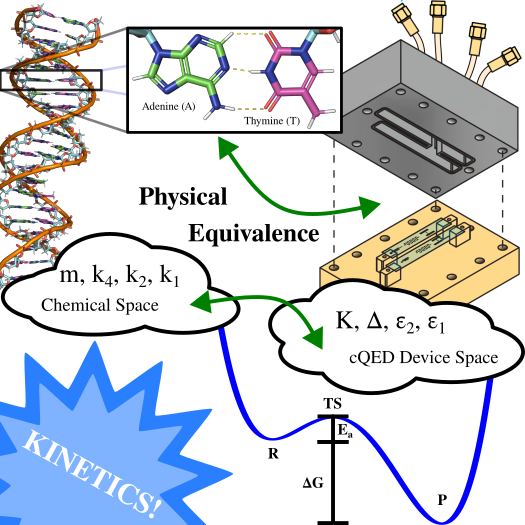}
\end{tocentry}

\begin{abstract} 
Chemical reactions are commonly described by the reactive flux transferring population from reactants to products across a double-well free energy barrier. Dynamics often involves barrier recrossing and quantum effects like tunneling, zero-point energy motion and interference, which traditional rate theories, such as transition-state theory, do not consider. In this study, we investigate the feasibility of simulating reaction dynamics using a parametrically driven bosonic superconducting Kerr-cat device. This approach provides control over parameters defining the double-well free energy profile, as well as external factors like temperature and the coupling strength between the reaction coordinate and the thermal bath of non-reactive degrees of freedom. We demonstrate the effectiveness of this protocol by showing that the dynamics of proton transfer reactions in prototypical benchmark model systems, such as hydrogen bonded dimers of malonaldehyde and DNA base pairs, could be accurately simulated on currently accessible Kerr-cat devices. 
\end{abstract}

% \section{Introduction} 
Computational modeling of reaction dynamics offers insights into the time scales and mechanisms of molecular transformations in chemical reactions, revealing the factors that determine the reaction rates and efficiencies. Most chemical reactions are multi-step processes, with each step described by the reactive flux transferring across a barrier of a double-well potential energy surface along the reaction coordinate. However, simulating these elementary steps can be challenging, especially when barrier recrossing and significant quantum effects are involved, such as tunneling, zero-point energy, and interference. These factors lead to a complex interplay between coherence and dissipative dynamics, which are not accounted for by traditional rate theories, such as transition-state theory.~\cite{eyring_activated_1935, truhlar_current_1996, nitzan_chemical_2006,pechukas_transition_1981, laidler_development_1983, pollak_reaction_2005}

Recent advances in quantum engineering have generated interest in developing quantum devices to simulate the quantum dynamics of atoms, molecules and condensed phase systems.
\cite{altman_quantum_2021, georgescu_quantum_2014} Such analog quantum simulators can offer significant hardware efficiency advantage over general-purpose quantum computers, especially when the device Hamiltonian can be efficiently mapped onto the molecular system of interest. In this study, we explore the feasibility of using a Superconducting Nonlinear Asymmetric Inductive eLement (SNAIL) transmon~\cite{frattini20173} to simulate quantum dynamics of elementary chemical reactions. The SNAIL device develops a double-well structure when operated under a continuous drive with frequency close to twice the SNAIL transmon resonance. This system is experimentally realizable in a superconducting quantum circuit and is referred to as Kerr-cat device.~\cite{Frattini2022} Its controllable parameters can be adjusted to model the asymmetric double-well free energy profiles of various molecular systems across a wide range of external conditions, including temperature and coupling strength between the system and the thermal bath of non-reactive degrees of freedom.
Here, we explore the capabilities of SNAIL devices by simulating the chemical dynamics of prototypical proton transfer reactions in hydrogen-bonded complexes, using the SNAIL Hamiltonian parametrized for malonaldehyde dimers and DNA base pairs, as could be experimentally implemented today on currently available platforms. These simulations are currently unattainable with noisy intermediate-scale quantum (NISQ) computers, including state-of-the-art conventional superconducting quantum computers, due to their circuit depth limitations. In contrast, we show that the SNAIL device could accurately capture the dynamics of these chemical reactions, effectively accounting for the delicate interplay between tunneling, zero-point energy, resonance, interference, and dissipative effects. 

In transition-state theory (TST),\cite{eyring_activated_1935, truhlar_current_1996, nitzan_chemical_2006}
the reaction rate constant is given by the Eyring-Polanyi equation: $k_{TST}=\frac{k_B T}{h} e^{-\Delta G^{\ddagger}/k_BT}$. Here, $\Delta G^{\ddagger}$ is the Gibbs free energy of activation (the barrier height as measured from the minimum of the reactant well to the transition state configuration that corresponds to the barrier top), $k_B$ is the Boltzmann constant, $T$ is the absolute temperature, and $h$ is the Planck constant. The Gibbs free energy of activation can be divided into enthalpy and entropy contributions: $\Delta G^{\ddagger} = \Delta H^{\ddagger}-T \Delta S^{\ddagger}$, where $\Delta H^{\ddagger}$ is the activation enthalpy and $\Delta S^{\ddagger}$ is the activation entropy. Identifying $\Delta H^{\ddagger}$ as the activation energy ($E_a$), the TST rate constant can be expressed by Arrhenius' law, 
$k_{TST}=A~e^{-E_a/k_BT}$, with $A=\frac{k_B T}{h} e^{\Delta S^{\ddagger}/k_B}$ the rate constant for the barrierless case (i.e., when $\Delta H^{\ddagger}=E_a=0$). 

Deviations from TST and the Arrhenius law occur when the assumptions of these theories become invalid.\cite{yamamoto_quantum_1960,
chandler_statistical_1978} Recrossing events and quantum phenomena like tunneling, quantum interference, and zero-point energy effects can lead to deviations from TST by introducing reactive pathways beyond classical barrier crossing.\cite{geva_quantum-mechanical_2001} 
Some of those deviations can be approximately accounted for by the transmission coefficient $\kappa$, modifying the reaction rate constant to $k = \kappa k_{TST}$. This approach assumes that the reaction dynamics follows rate kinetics with a single well-defined rate constant. 
Another breakdown of TST occurs when the concept of a rate constant becomes invalid, such as when the barrier height is comparable to $k_B T$ 
or when the coupling between the reaction coordinate and other non-reactive molecular degrees of freedom (DOF) is weak.\cite{vazquez_entropic_2014} These forms of TST breakdown often stem from the inherently quantum-mechanical nature of chemical dynamics.\cite{miller_quantum_1974, miller_quantum_1983, miller_direct_1998}

A prime example of an elementary chemical reaction of fundamental biological importance, which is often modeled using a double-well potential energy surface, is the adenine-thymine proton transfer reaction in DNA.~\cite{godbeer2015modelling,soley2021iterative,soley2021functional} The free energy profile for this reaction varies with physiological conditions. Under normal cell conditions, the profile favors the hydrogen-bonded adenine-thymine complex form. However, during cell replication, this hydrogen bond must be broken to allow the DNA strands to duplicate. Proton transfer during this process has been suggested to cause spontaneous mutations due to bases occasionally adopting less likely tautomeric forms.~\cite{watson1993genetical} Therefore, simulating the dynamics of proton transfer in the adenine-thymine complex across a broad range of double-well free energy profiles can provide insights into the potential influence of quantum effects on the interconversion between tautomeric forms of the bases.

The traditional method for testing and validating chemical rate theories, including both TST/Arrhenius and post-TST/non-Arrhenius approaches, has relied on extensive experimental measurements of chemical dynamics across a diverse range of molecular systems and external conditions to cover both chemical space and parameters influencing chemical dynamics, such as temperature and the interactions between the reaction coordinate and the non-reactive DOF. However, this approach is highly challenging and labor-intensive. This is because experimentally monitoring reactant and product populations in real time may prove difficult, and changing from one chemical system to another often involves altering multiple parameters with complex and sometimes opposite effects. A notable example of the difficulty of this traditional approach is the 30-year delay between the theoretical prediction of the inverted region in the Marcus rate theory for electron transfer reactions and its experimental validation.~\cite{closs_intramolecular_1988, miller_intramolecular_1984, marcus_electron_1993}

A promising alternative for experimentally testing and validating chemical rate theories has emerged with the advent of controllable and highly tunable fully quantum-mechanical platforms. These platforms can allow the exploration of quantum dynamics in various complex model systems across a broad range of parameter space.\cite{georgescu_quantum_2014, altman_quantum_2021,dutta2024simulatingchemistrybosonicquantum}
Considering that chemical dynamics is inherently quantum mechanical in nature and that the most under-explored regimes of chemical dynamics exhibit significant quantum effects, these platforms can offer valuable simulation tools to investigate chemical reactivity under conditions that could be challenging for spectroscopic methods applied to molecular systems.\cite{altman_quantum_2021}

Recent studies that exemplify this innovative approach of using quantum devices to explore chemical reactivityinclude the use of an ion trap platform as an analog simulator of the chemical dynamics underlying redox electron transfer reactions.~\cite{schlawin21,so2024trappedionquantumsimulationelectron}
Marcus theory, which describes the electron transfer rate constant with a double-well model, is analogous to TST for non-redox chemical reactions. It employs a TST-like argument with an Arrhenius-type expression for the rate constant.\cite{nitzan_chemical_2006} In Marcus theory, the transition state is the molecular configuration at the crossing point between the diabatic free energy profiles of the donor (reactant) and acceptor (product) states along the reaction coordinate associated with the reorganization of the nuclear DOF upon electron transfer. Like TST, Marcus theory treats the activation to the transition state as a classical process and assumes weak electronic coupling between the donor and acceptor states.  Schlawin et al. demonstrated that an ion-trap device could reproduce the predictions of Marcus theory, including the inverted regime.~\cite{schlawin21} They also showed how this device could explore deviations from Marcus theory because of quantum (low-temperature) and strong electronic coupling effects. This enables the study of unconventional electron transfer regimes that are challenging to capture using traditional methods. 

In this paper, we propose a novel strategy for simulating the dynamics of elementary reactions using a quantum platform. We explore the superconducting circuit quantum electrodynamics (cQED) Kerr-cat device.~\cite{frattini_squeezed_2022, venkatraman_driven_2023, Reynoso2023,
chavez-carlos_spectral_2023, Iachello2023, GarciaMata2024, Chavez2024Chaos,dutta2024simulatingchemistrybosonicquantum}
We begin by introducing the device and highlighting the features that make it ideally suited for analog simulations of chemical dynamics. Next, we consider examples of molecular systems to demonstrate the Kerr-cat device capabilities as applied to simulating proton transfer dynamics. 

\begin{figure}[ht!]
    \centering
    \includegraphics[width=0.9\textwidth]{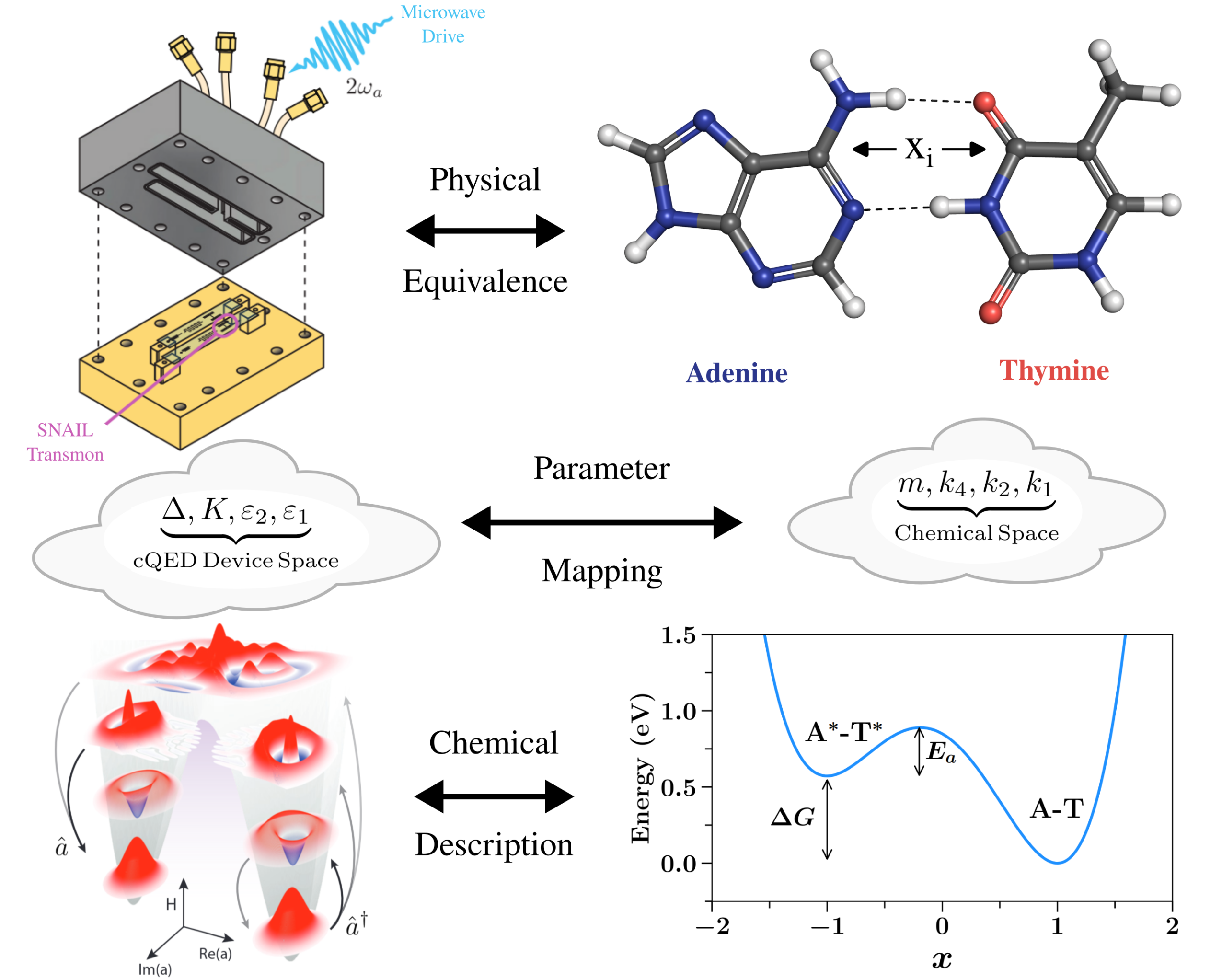}
    \caption{Schematic of Bosonic cQED Device for Quantum Dynamics Simulations in Hydrogen-Bonded Complexes. Adapted from Ref.~\citenum{dutta2024simulatingchemistrybosonicquantum}. Copyright (2024) American Chemical Society. The device is a half-aluminum, half-copper cavity resonator containing two sapphire chips with a SNAIL-transmon, readout resonator, and Purcell filter (top left).\cite{Albornoz2024} A strong microwave drive at twice the resonance frequency converts the SNAIL-transmon Hamiltonian into the bi-stable Kerr-cat parametric oscillator Hamiltonian for analog simulations of quantum dynamics in molecular systems, such as the adenine-thymine dimer (top right). Parameters $\Delta, K, \epsilon_1, \epsilon_2$ (middle left) are adjusted to map the effective Hamiltonian $\hat{H}_{KC}$ to the double-well Hamiltonian modeling the molecular system (middle right). The bottom left shows the Wigner transform phase space representation of the device quantum states, which are analogous to the states in molecular the double-well potential in the bottom right.
    }
    \label{fig:fig1}
\end{figure}

% \section{Dissipative Dynamics of the cQED Kerr-cat Device} 
We begin by considering the effective Hamiltonian of the cQED Kerr-cat device, which constitutes of an arrangement of a few Josephson junctions schematically shown in Fig.~\ref{fig:fig1}:~\cite{frattini_squeezed_2022,NickFrattini_thesis}
\begin{equation}
    \hat{H}_{KC}
    = \Delta \hat{a}^{\dagger} \hat{a} - K 
    (\hat{a}^{\dagger})^2 (\hat{a})^2 + \epsilon _2 (\hat{a}^2 + \hat{a}^{\dagger 
    2}) + \epsilon _1 (\hat{a} + \hat{a}^{\dagger}).
    \label{eq:kerr-cat}
\end{equation}
Here, $\hadag$ and $\ha$ are the ladder operators for the device bosonic mode, satisfying the usual bosonic commutator relation $[\ha, \hadag]=1$. The Hamiltonian $\hat{H}_{KC}$ includes adjustable parameters, namely the Kerr non-linearity, $K$, the detuning parameter, $\Delta$, and the drive coefficients, $\epsilon_1$ and  $\epsilon_2$. These parameters can be experimentally altered by adjusting the magnetic flux and the amplitudes and frequencies of the microwave drives %of 
within the quantum device.\cite{Albornoz2024}

Following Venkatraman {\em et al.},~\cite{venkatraman_driven_2023} we assume that the noisy dynamics of the device can be described by the following Lindblad equation: 
\begin{align}\label{eq:Lindbladian_aadag}
    \dhr = -\dfrac{i}{\hbar} [\hH_{KC}, \hr] 
            + \kappa~(1 + \nth) \lrp{\ha \hr \hadag 
                                                 - \dfrac{1}{2} \lrB{\hadag \ha, \hr}}
                              + \kappa~\nth \lrp{\hadag \hr \ha 
                                                 - \dfrac{1}{2} \lrB{\ha \hadag, \hr}},
\end{align}
where $\kappa$ sets the photon loss rate corresponding to an effective dissipation rate due to coupling of the system with a surrounding environment  and $\nth$ represents the average thermal photon population, a quantity determined by the temperature. Like $\{ K, \Delta, \epsilon_1, \epsilon_2 \} $, the parameters $\kappa$ and $\nth$ are experimentally tunable and determine the noise associated with the experimental device.~\cite{ding2024quantumcontroloscillatorkerrcat}

\begin{figure}[ht!]
    \centering
    \includegraphics[width=0.8\textwidth]{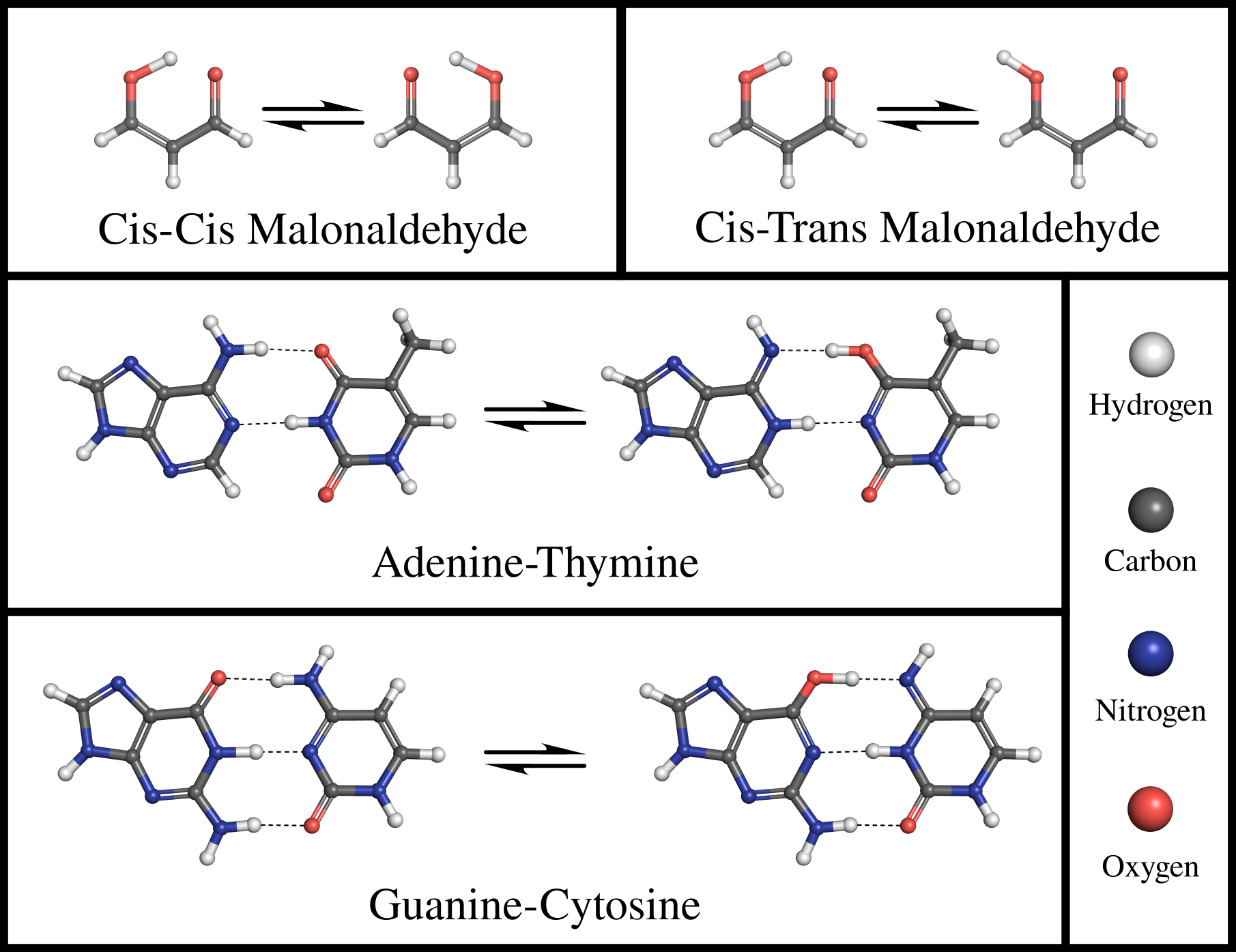}
    \caption{Hydrogen bonded complexes studied to analyze intramolecular proton transfer in cis-cis and cis-trans malonaldehyde (top), intramolecular proton transfer in adenine-thymine (purine and pyrimidine) (middle), and guanine-cytosine (purine and pyrimidine) base pairs.
    }
    \label{fig:studied_systems}
\end{figure}

% \section{Mapping of the Molecular Model Hamiltonian} 
Consider also the model Hamiltonian for systems with a double-well potential energy surface given by
\begin{align}
    \hat{H}_{DW}    & = \frac{\hat{p}^2}{2m} +  k_4 \hat{x}^4 - k_2 
    \hat{x}^2+k_1 \hat{x}, 
    \label{eq:dwps}
\end{align}
which is commonly used to simulate hydrogen bonded complexes.~\cite{godbeer2015modelling} The position and momentum operators of the reaction coordinate, $\hat{x}$ and $\hat{p}$, satisfy the commutation relation $[\hat{x} , \hat{p} ] = i \hbar$. In the equation above, $m$ represents the effective mass associated with motion along the reaction coordinate. The parameters 
$\{ k_1, k_2, k_4 \}$ are positive and real, typically determined by fitting the {\em ab initio} potential energy surface. This fitting process ensures the accurate representation of the barrier height, the curvature of the surface at the reactant and product wells, and the relative stability of reactants and products as parametrized by $k_1$. Table \ref{tab:fit_params} lists the parameters $\{ k_1,k_2,k_4\}$ used in this study for numerical simulations of proton transfer dynamics (with $m=1836$ amu) in the four model systems illustrated in Fig.~\ref{fig:studied_systems}.

\begin{table}[ht]
    \centering
    \begin{tabular}{cccc}
        \toprule
        System & $k_4$ [$\mathrm{E_h/a_0^4}$] & $k_2$
        [$\mathrm{E_h/a_0^2}$] & $k_1$ [$\mathrm{E_h/a_0}$] \\ \midrule
        Adenine-Thymine (DNA) \cite{AT_potential} & $1.4\times 10^{-3}$ & $1.08\times 10^{-2}$ & $5.2\times 10^{-3}$  \\ 
        Guanine-Cytosine (DNA) \cite{slocombe_open_2022} & $7.7\times 10^{-4}$ & $6.9 \times 10^{-3}$ & $4.5\times 10^{-3}$ \\ 
        Malonaldehyde (cis-trans) \cite{ghosh_dynamics_2012, ghosh_optimised_2015} & $9.4\times 10^{-5}$ & $3.0\times 10^{-3}$ & $2.9\times 10^{-3}$ \\ 
        Malonaldehyde  (cis-cis) \cite{ghosh_dynamics_2012, ghosh_optimised_2015} & $7.1\times 10^{-4}$ & $4.0\times 10^{-3}$ & 0 \\
        \bottomrule
    \end{tabular}
    \caption{
    Parameters used to simulate double-well potentials for proton transfer in molecular systems, according to the equation $V = k_4 x^4 - k_2 x^2
    + k_1 x$.}
    \label{tab:fit_params}
\end{table}

The next step is to map the Kerr-cat Hamiltonian, introduced in Eq. (\ref{eq:kerr-cat}), onto the model Hamiltonian of the molecular system presented in Eq. (\ref{eq:dwps}). This requires mapping the photonic operators, $\{ \hat{a} , \hat{a}^\dagger\}$, onto the reaction coordinate operators, $\{ \hat{x} , \hat{p} \}$. To ensure the correct dynamics, this mapping must preserve the commutation relations: $[\hat{x} , \hat{p} ] = i \hbar$ and $[\hat{a} , \hat{a}^\dagger ] = \hat{1}$. A mapping that satisfies these conditions is given by:
\begin{align}\label{eq:aaxp_mapping}
    \hat{a} = \dfrac{1}{\sqrt{2}} \lrp{
    \dfrac{1}{c}\hx + \dfrac{i c}{\hbar}\hp}, \qquad
    \hadag =\dfrac{1}{\sqrt{2}}\lrp{\dfrac{1}{c}\hx - \dfrac{i c}{\hbar}\hp},
\end{align}
where $c$ is an arbitrary parameter  with the same units as $\hat{x}$.
Thus, the mapping of $\{ \hat{a} , \hat{a}^\dagger\}$ onto $\{ \hat{x} , \hat{p} \}$ is not unique. This flexibility in choosing $c$ plays a crucial role in mapping the Kerr-cat Hamiltonian in Eq.~(\ref{eq:kerr-cat}) onto the chemical double-well Hamiltonian in Eq. (\ref{eq:dwps}).
It should be noted that the value of $c$ can be equivalent to the zero-point spread, $\sqrt{\hbar Z}$, if we define $\hx$ and $\hp$ as in reference~\citenum{venkatraman_driven_2023}.

Substituting the expressions for $\hat{a}$ and $\hat{a}^\dagger$ in terms of $\hat{x}$ and $\hat{p}$ from Eq.~(\ref{eq:aaxp_mapping}) into Eq.~(\ref{eq:kerr-cat}), we can recast the {\em negative} of the Kerr-cat Hamiltonian in terms of $\hat{x}$ and $\hat{p}$, omitting constant terms that do not impact the dynamics:
\begin{align}
    -\hat{H}_{KC} &= \frac{c^2}{\hbar^2} \left( \epsilon_2 - K 
    -\Delta/2\right)\hat{p}^2 
    +\frac{K}{4c^4} \hat{x}^4 
    - \frac{1}{c^2}\left( 
\epsilon_2+K+\Delta/2\right)\hat{x}^2 
    -\frac{\epsilon_1 
    \sqrt{2}}{c}\hat{x}
    \notag 
    \label{eq:kcps}\\ &+ \frac{Kc^4}{4\hbar^4}\hat{p}^4 
    +\frac{K}{4\hbar^2}\lrp{\hat{x}^2\hat{p}^2+\hat{p}^2\hat{x}^2}.
\end{align}
Comparing Eq.~(\ref{eq:kcps}) with Eq.~(\ref{eq:dwps}), we see that while $-\hat{H}_{KC}$ includes $\hat{p}^2$, $\hat{x}^4$, $\hat{x}^2$ and $\hat{x}$ terms that can be mapped to the corresponding terms in the chemical double-well Hamiltonian (Eq.~(\ref{eq:dwps})), it also contains additional terms ($\hat{p}^4$, $\hat{x}^2\hat{p}^2$ and $\hat{p}^2\hat{x}^2$) that are absent in Eq.~(\ref{eq:dwps}). 

We now map the parameters $\{ \Delta,K,\epsilon _2,\epsilon _1 \}$ onto $\{ m, k_1 , k_2, k_4 \}$ by equating the coefficients of the $\hat{p}^2$, $\hat{x}^4$, $\hat{x}^2$ and $\hat{x}$ terms in Eqs.~(\ref{eq:dwps}) and (\ref{eq:kcps}). This leads to the following mapping relations:
\begin{align}
    K &=  4c^4 k_4,
        \label{eq:K_param} \\
    \epsilon_2 &= \dfrac{\hbar^2}{4 c^2 m} + \dfrac{c^2 k_2}{2},
        \label{eq:e2_param} \\
    \Delta &= - \dfrac{\hbar^2}{2 c^2 m} + c^2 k_2 - 8 c^4 k_4,
        \label{eq:delta_param} \\
    \epsilon_1 &= -\dfrac{c k_1}{\sqrt{2}}.
        \label{eq:e1_param}
\end{align}
Clearly, according to Eqs.~(\ref{eq:K_param})-(\ref{eq:e1_param}), the values of $\{ \Delta,K,\epsilon _2,\epsilon _1 \}$ depend on the value of $c$. We then utilize the flexibility in choosing the value of $c$ to minimize the effect of the additional terms $\hat{p}^4$, $\hat{x}^2\hat{p}^2$ and $\hat{x}^2\hat{p}^2$ in Eq.~(\ref{eq:dwps}). As demonstrated in the SI, this requires choosing a value of $c$ small enough to satisfy the following inequality:
\begin{align}
    \dfrac{\hbar^2}{ m k_2 c^4} \gg 1.
    \label{eq:ineq1}
\end{align}

Next, we compare the energy levels of the Kerr-cat and double-well Hamiltonians, given by Eq.~(\ref{eq:kcps}) and Eq.~(\ref{eq:dwps}), respectively. In general, these energy levels are expected to be {\em different} due to the additional terms $\hat{p}^4$, $\hat{x}^2 \hat{p}^2$ and $\hat{p}^2 \hat{x}^2$ of the Kerr-cat Hamiltonian that are missing in the double-well Hamiltonian. However, the deviations between the energy levels of both Hamiltonians are expected to decrease as $c$ becomes smaller [see Eq.~(\ref{eq:ineq1}) and Fig.~\ref{fig:hamiltonian_analysis}, which is explained below].

To establish an acceptable value of $c$, we set the tolerance for deviations between the energy levels of the Kerr-cat and double-well Hamiltonians at 1.5 m$\mathrm{E_{h}}~\text{= 0.941 kcal mol}^{-1}$, a standard measure of chemical accuracy. It is crucial to note that accurate descriptions of barrier crossing dynamics require this level of accuracy not only for the ground state but also for energy levels up to the top of the barrier. 

\begin{figure}[ht!]
    \centering
    \makebox[\textwidth][c]{
        \includegraphics[width=1.0\textwidth]{
        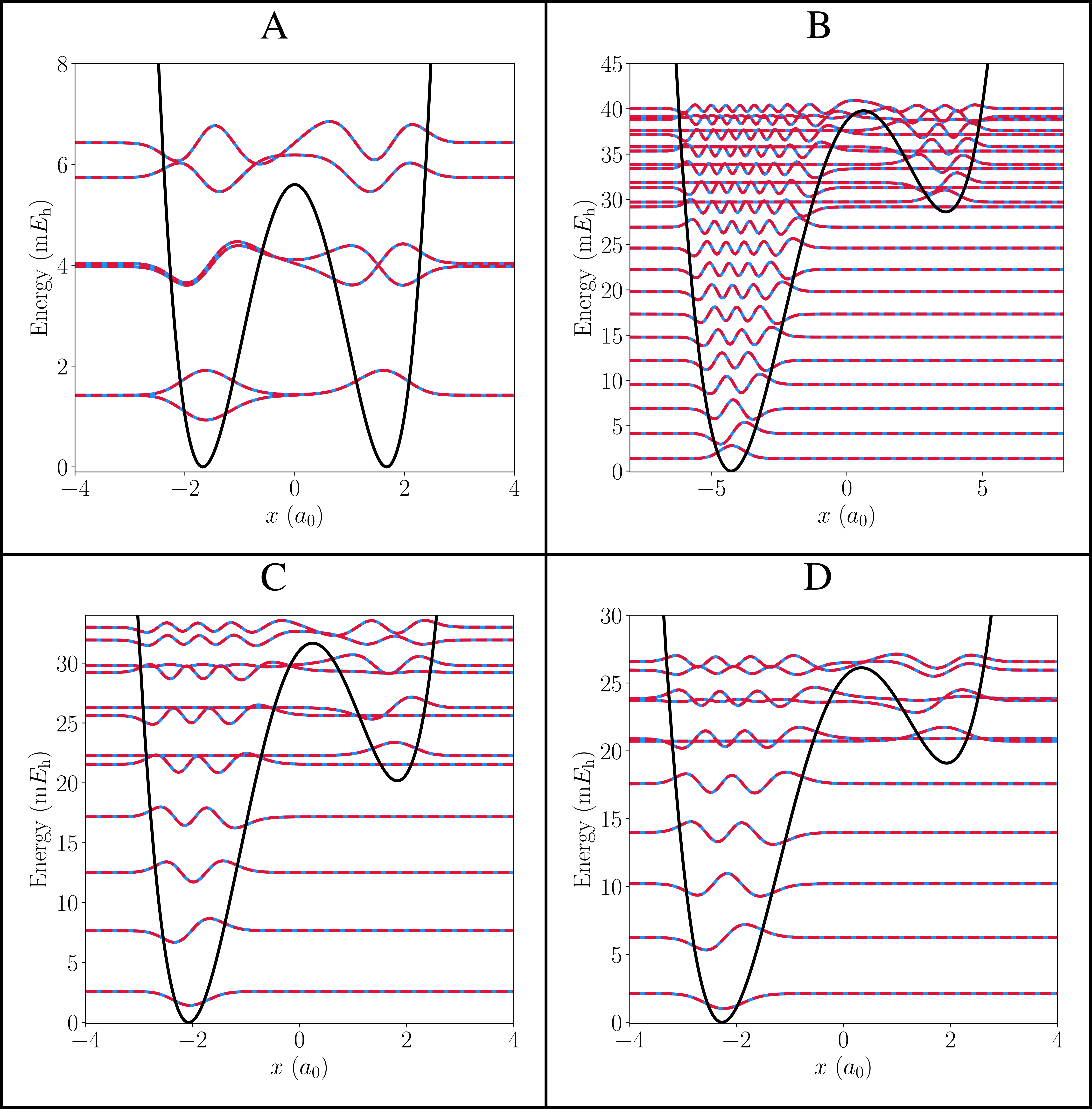}
    }
    \caption{Eigenstates obtained by diagonalization of the model Hamiltonians for cis-cis malonaldehyde (A), cis-trans malonaldehyde (B), adenine-thymine (C), and guanine-cytosine (D), based on either the chemical double-well Hamiltonian (blue) or the Kerr-cat Hamiltonian (red) with $c=0.1 \, \mathrm{a_0}$. Eigenstates are shown together with the potential energy surface as a function of the reaction coordinate `$x$'. 
        }
    \label{fig:ham_eigenstates}
\end{figure}

\begin{figure}[ht!]
    \centering
    \makebox[+0pt][c]{
        \includegraphics[width=1.0\textwidth]{
        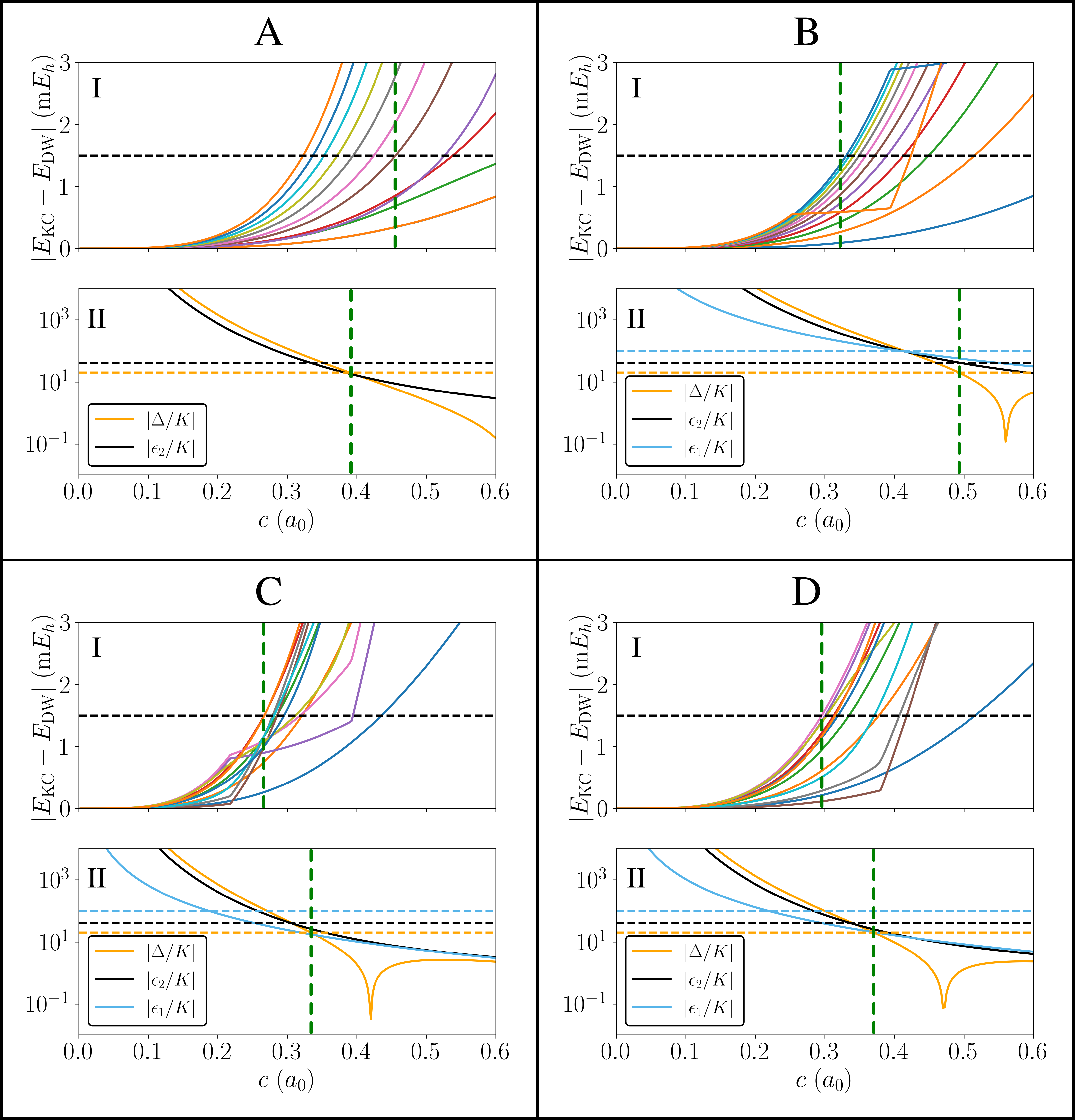}
    }
    \caption{Absolute energy differences for the lower lying eigenstates of the chemical double-well and Kerr-cat Hamiltonian (panels I) and cQED device parameters (panels II) as a function of the scaling parameter $c$ for cis-cis malonaldehyde (A), cis-trans malonaldehyde (B), adenine-thymine (C) and guanine-cytosine (D). The horizontal black dashed line in panels I indicates the threshold of chemical accuracy at 1.5 m$\mathrm{E _h}$. The horizontal dashed lines in panels II indicate the maximum values of the parameters available in current cQED devices. The vertical dashed line indicates the maximum value of $c$ recommended for accurate dynamics  simulations (panels I) and minimum value of $c$ ensuring experimental parameters available in existing cQED Kerr-cat platforms (panels II). \cite{Albornoz2024}
    }
    \label{fig:hamiltonian_analysis}
\end{figure}

For example, the number of states required to capture accurate dynamics in the molecular model systems analyzed in this work is as follows: 6 for cis-cis malonaldehyde (Fig.~\ref{fig:ham_eigenstates}.A), 
24 for cis-trans malonaldehyde (Fig.~\ref{fig:ham_eigenstates}.B), 12 for adenine-thymine (Fig.~\ref{fig:ham_eigenstates}.C), and 14 for guanine-cytosine (Fig.~\ref{fig:ham_eigenstates}.D). Ensuring that the energy levels of the Kerr-cat device, $E_{KC}$, and chemical double-well Hamiltonians, $E_{DW}$, are within chemical accuracy of each other ($E_{KC}-E_{DW}$ is below the horizontal dashed line in the panels I of Fig.~\ref{fig:hamiltonian_analysis}) also results in excellent agreement between their corresponding eigenfunctions, as shown in Fig.~\ref{fig:ham_eigenstates}.

Using the cQED device to encode $\hdw$ requires a finite, non-zero value for the Kerr non-linearity $K$, so the units of $\Delta$, $\epsilon_2$ and $\epsilon_1$ are expressed in terms of $K$. For vanishingly small values
of $c$, the ratios $\frac{\Delta}{K}$,$\frac{\epsilon_2}{K}$, and $\frac{\epsilon_1}{K}$ become very large (panels II of Fig. \ref{fig:hamiltonian_analysis}), so they are experimentally unfeasible with current cQED platforms.~\cite{Albornoz2024} However, there exists a range of values of $c$ for which the parameters are experimentally accessible, while maintaining a useful degree of accuracy for the energies and stationary states. In particular, the Hamiltonian for the cis-cis malonaldehyde is both experimentally accessible and meets the chemical accuracy criterion for stationary states and energies (see Fig. \ref{fig:hamiltonian_analysis}A). Chemical systems with asymmetric free energy profiles
pose greater challenges due to the higher number of eigenstates required for addressing kinetic questions. Expanding the range of experimental parameters could enable an accurate simulation of these more challenging double-well problems, although it may also risk a breakdown of the effective Hamiltonian approximation.\cite{Albornoz2024}

% \section{Device Dynamics versus Chemical Dynamics}
Consistency between the energy levels of the $\hat{H}_{KC}$ and $\hat{H}_{DW}$ Hamiltonians is necessary, but not sufficient to ensure that analog simulations on the Kerr-cat device accurately capture the chemical dynamics. 
This is because the {\em reaction rate} is also influenced by the coupling of the reaction coordinate to the thermal bath of {\em nonreactive} DOF. Consequently, the actual reaction rate constant depends not only on the eigenvalues and eigenfunctions of the Hamiltonian but also on the temperature and dissipation rate of the bath. Therefore, being able to simulate chemical dynamics on the Kerr-cat device requires that the dynamics of the device be dissipative (noisy), which is indeed the case. The noisy dynamics of Kerr-cat device are described by the Lindblad equation in Eq.~(\ref{eq:Lindbladian_aadag})
with adjustable photon loss rate constant $\kappa$ and
temperature set by $n_{th}$. 

In what follows, we consider a chemical system that experiences the same type of dissipation as the Kerr-cat device. This dissipative model is useful as it results in chemical dynamics characterized by a reaction rate constant. However, other dissipation models could also be useful, as long as they capture the dissipation of the chemical system typically determined by the nature of its chemical environment and the way the reaction
coordinate is coupled to the environment ({\em e.g.}, the surrounding can correspond to liquid solution, biological environment, or solid-state environment). Hence, characterizing the dissipation of  specific chemical systems and replicating it on a Kerr-cat device remains a subject of future studies that will focus on bath engineering.~\cite{murch2012cavity,kitzman2023phononic} 
It is noteworthy that Markovian and exotic dissipative channels can be engineered to a great extent in circuit QED  \cite{Shen2017_PRB}. The strength of dissipation in the squeezed Kerr oscillator experiment of Frattini and coworkers\cite{frattini_squeezed_2022} is, for example, readily tunable by the amplitude of the microwave readout drive, and exotic forms of dissipation have been already demonstrated \cite{sivak2023real,leghtas2015confining}. 

Assuming that the chemical system undergoes the same type of dissipation as the Kerr-cat device, we use Eq.~(\ref{eq:aaxp_mapping}) to express the dissipator in Eq.~(\ref{eq:Lindbladian_aadag}) in terms of the operators $\hat{x}$ and $\hat{p}$, which form the basis of the chemical double-well Hamiltonian in Eq.~(\ref{eq:dwps}). This results in the following Lindblad equation for the chemical system: 
\begin{align}\label{eq:Lindbladian_xp}
    \dhr &= -\dfrac{i}{\hbar} [\hH_{DW}, \hr] 
         + \dfrac{\kappa \lrp{1+2\nth}}{4} 
         \lrb{ 
            \dfrac{1}{c^2} \lrp{\lrb{\hx \hr, \hx} + \lrb{\hx, \hr \hx}}
            + \dfrac{c^2}{\hbar^2} \lrp{\lrb{\hp \hr, \hp} + \lrb{\hp, \hr \hp}}
            } \nnn
        &- \dfrac{i \kappa}{4 \hbar} \lrp{ \lrb{\hx \hr, \hp} + \lrb{\hx, \hr \hp}   - \lrb{\hp \hr, \hx} - \lrb{\hp, \hr \hx}
        }.
\end{align}

The complete methodology for the numerical simulation of the open quantum dynamics is outlined in the SI. Here, we clarify that the initial state for dynamics propagation is localized in the reactant well, which is the higher-energy well of the above-mentioned chemical reactions. To generate this state, we diagonalize the system Hamiltonian and select the first eigenstate with more than 50\% density in the reactant well. We then apply a sigmoidal filter function to remove the excess density outside of it (see SI for implementation details). This method creates a localized state and avoids the need to identify the critical points of each potential energy surface. By integrating Eq. (\ref{eq:Lindbladian_xp}) with an initial state $\hat{\rho}(0)$ localized in the reactant well, we obtain the time-evolved state $\hat{\rho} (t)$, which can be used to compute the product population at time $t$, as follows: 
\begin{equation}
P_P (t) = \mathrm{Tr}\{\hr(t) \hat{\Theta}_X\},
\label{eq:prod_pop}
\end{equation}
where $\hat{\Theta}_X$ is the Heaviside function:
\begin{equation}
\langle x | \hat{\Theta}_X | x' \rangle = \left\{ 
\begin{array}{cc}
0 & \mbox{for } x<0,\\
\delta (x-x') & \mbox{for } x>0. 
\end{array}
\right.
\label{eq:heaviside}
\end{equation}

\begin{figure}[ht!]
    \centering
    \makebox[\textwidth][c]{
        \includegraphics[width=0.9\textwidth]{
        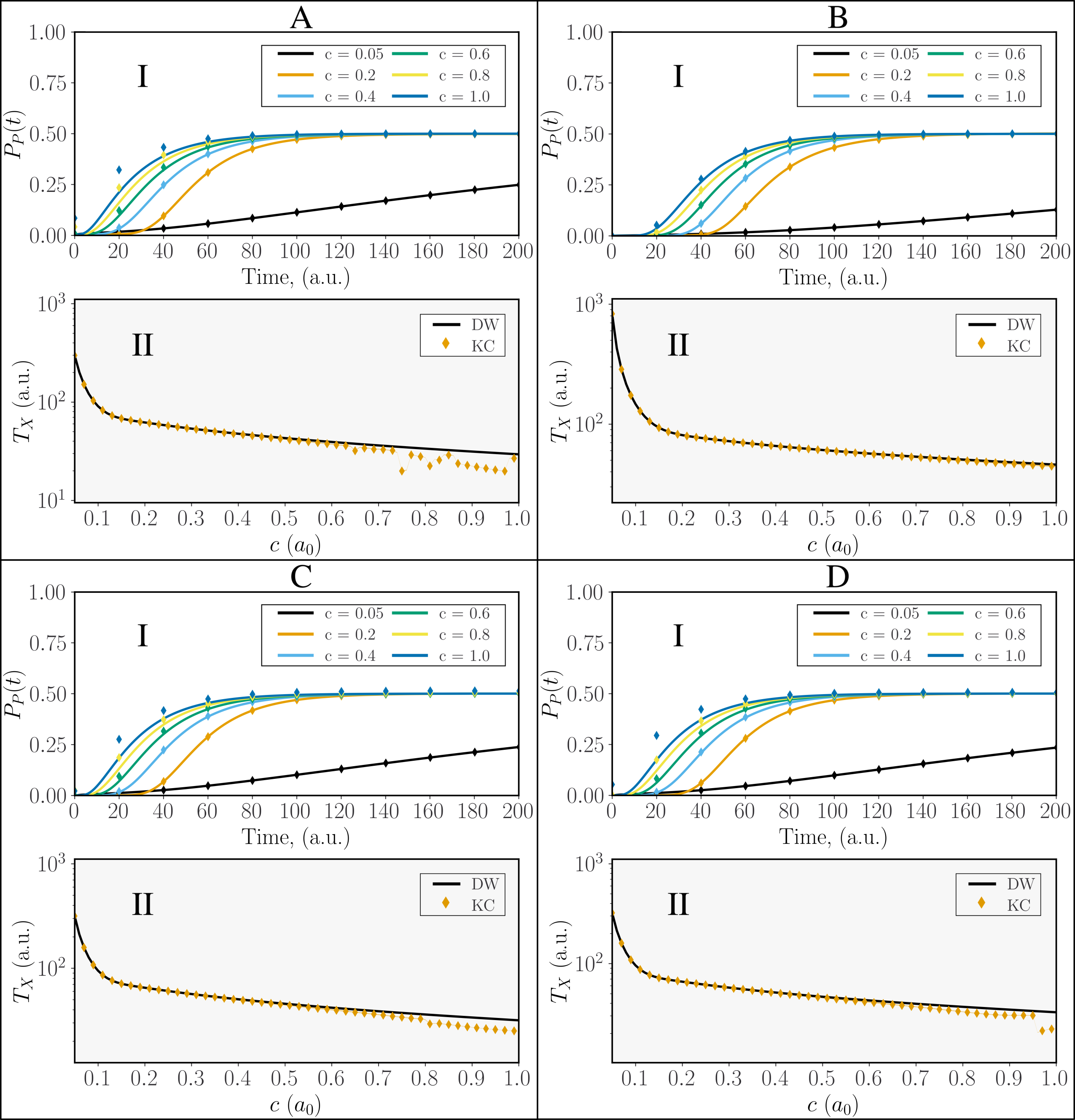}
    }
    \caption{Comparison of observables obtained with $\hat{H}_{DW}$ (solid lines) and $\hat{H}_{KC}$ (diamonds) as a function of $c$, using $\kappa=0.1$ and $\nth=0.1$, for cis-cis malonaldehyde (A), cis-trans malonaldehyde (B), adenine-thymine (C) and guanine-cytosine (D). The time evolution of the product population is shown in the top subpanels (I), while the corresponding inverse reaction rate constants are shown in the bottom subpanels (II).
    }
    \label{fig:lindbladian_dynamics}
\end{figure}

The reactant-to-product reaction rate constant, $k = 1/T_X$, is obtained by fitting $P_P (t)$ to an exponential decay. The top subpanels of Fig.~\ref{fig:lindbladian_dynamics} show the time evolution of $P_P (t)$ for the four proton transfer reactions -- cis
malonaldehyde, cis-trans malonaldehyde, adenine-thymine, and guanine-cytosine -- obtained by solving Eq.~(\ref{eq:Lindbladian_xp}) for different values of $c$ (solid lines). These top subpanels also include results where the chemical double-well Hamiltonian in Eq.~(\ref{eq:Lindbladian_xp}), $\hat{H}_{DW}$, is replaced by the corresponding device Hamiltonian $-\hat{H}_{KC}$ with additional terms $\hat{p}^4$, 
$\hat{x}^2 \hat{p}^2$ and $\hat{p}^2 \hat{x}^2$ (diamonds). The bottom subpanels of Fig.~\ref{fig:lindbladian_dynamics} show the dependence of the inverse reaction rate constant, $T_X$, on $c$ for both $\hat{H}_{DW}$ (solid line) and $-\hat{H}_{KC}$ (diamonds). 

A close examination of the results in Fig.~\ref{fig:lindbladian_dynamics} reveals that the dynamics of both the chemical system and the Kerr-cat device align well, with the rate kinetics described by a rate constant. Additionally, the reaction rate constant shows significantly less sensitivity to the value of $c$ compared to the energy levels and eigenfunctions of $\hat{H}_{DW}$ and $\hat{H}_{KC}$. Specifically, the rate constant predicted by the device matches that predicted for the chemical system at values of $c$ as high as 0.4 $\mathrm{a_0}$, much higher than the value of $c=0.1 \, \mathrm{a_0}$ required to match energy levels and eigenfunctions near the top of the barrier. This is likely due to the effect of dissipation, making the overall dynamics less sensitive to small differences in the potential energy surface. 

The results in Fig.~\ref{fig:lindbladian_dynamics} were obtained with $\kappa=0.1$ and $\nth=0.1$, which are illustrative of the dissipation on the quantum device. However, the observed trends are insensitive to these specific values, as demonstrated by the rate constants obtained for various values of $\kappa$ and $\nth$ shown in Table~\ref{tab:rate_constants} for $c=0.1 \, \mathrm{a_0}$. As expected, $T_X$ increases ({\em i.e.}, the reaction slows down) when $\kappa$ and $\nth$ decrease. Nonetheless, the actual values of $T_X$ for the chemical 
double-well and the Kerr-cat device are consistent, regardless of the $\kappa$ and $\nth$ values. 

\begin{table}[ht!]
    \centering
    \begin{tabular}{c|cccc}
        \toprule
         Dissipation Constants $\lrp{\kappa, \nth}$ & $\lrp{0.1, 0.1}$ & $\lrp{0.1, 0.05}$ & $\lrp{0.025, 0.1}$ & $\lrp{0.025, 0.05}$ \\
         \midrule \midrule
         Cis-cis Malonaldehyde (KC)   & $91\pm1$  & $91\pm1$  & $303\pm4$  & $295\pm4$ \\
         Cis-cis Malonaldehyde (DW)   & $91\pm1$  & $91\pm1$  & $303\pm4$  & $295\pm4$ \\
         \midrule
         Cis-trans Malonaldehyde (KC) & $147\pm2$ & $142\pm2$ & $527\pm7$  & $499\pm6$ \\
         Cis-trans Malonaldehyde (DW) & $147\pm2$ & $142\pm2$ & $528\pm7$  & $500\pm6$ \\
         \midrule
         Adenine-Thymine (KC)         & $95\pm1$  & $94\pm1$ & $323\pm4$  & $314\pm4$ \\
         Adenine-Thymine (DW)         & $95\pm1$  & $94\pm1$ & $323\pm4$  & $314\pm4$ \\
         \midrule
         Guanine-Cytosine (KC)        & $96\pm1$  & $95\pm1$  & $325\pm4$  & $316\pm4$ \\
         Guanine-Cytosine (DW)        & $96\pm1$  & $95\pm1$  & $325\pm4$  & $316\pm4$ \\
         \bottomrule
    \end{tabular}
    \caption{Table of inverse reaction rate constants $T_X$ (in $\mathrm{\hbar/E_h}$) for each chemical system obtained with the $\hat{H}_{DW}$ and $\hat{H}_{KC}$ Hamiltonians, for various $\kappa$ and $\nth$ values with $c=0.1 \, \mathrm{a_0}$} 
    \label{tab:rate_constants}
\end{table}

% \section{Concluding Remarks} 
{\em Conclusions.--} In this paper, we proposed a novel approach to simulating chemical dynamics 
using a tunable Kerr-cat quantum device. This method allows for precise control over the parameters defining double-well potential energy surfaces, as well as external factors such as temperature and dissipation rates. We demonstrated the efficacy of this approach by applying it to proton transfer in four prototypical hydrogen-bonded model complexes, showing that the underlying chemical dynamics can be accurately simulated on a quantum device. 

Simulating chemical dynamics on a Kerr-cat device requires overcoming several 
challenges. A primary challenge, addressed in this paper, is mapping the chemical double-well Hamiltonian onto the Kerr-cat device Hamiltonian. This mapping is nontrivial, because the Kerr-cat device Hamiltonian includes additional terms ($\hat{p}^4$, $\hat{x}^2 \hat{p}^2$ and $\hat{p}^2 \hat{x}^2$), that are absent in the chemical Hamiltonian. We resolved this challenge by introducing a method that minimizes the impact of those additional terms by adjusting the parameter $c$ in the mapping of the photonic operators $\hat{a}$ and $\hat{a}^\dagger$ to the chemical operators $\hat{x}$ and $\hat{p}$. Specifically, we demonstrated that the energy levels and stationary states of the Kerr-cat and chemical Hamiltonians can be aligned (within chemical accuracy tolerance) by selecting a sufficiently small value of $c$. Furthermore, we found that reaction rate constants are even less sensitive to the value of $c$ than the energy levels and stationary states, making them easier to reproduce when simulating chemical dynamics on the Kerr-cat device.   

The approach proposed in this paper represents a significant first step towards enabling simulations of chemical dynamics on bosonic quantum simulators, which is beyond the capabilities of currently available NISQ computers. Remaining challenges include characterizing and engineering dissipation, and designing devices that can reliably emulate more complex free energy surfaces beyond the one-dimensional asymmetric double-well free energy surfaces considered in this paper. Ongoing work  addressing these challenges will be reported in future papers, paving the way for even more advanced simulations of quantum chemical dynamics on quantum devices.

\begin{acknowledgement}
The authors acknowledge support from the NSF Center for Quantum Dynamics on Modular Quantum Devices (CQD-MQD) under grant number 2124511. LFS and VSB acknowledge partial support from the National Science Foundation Engines Development Award: Advancing Quantum Technologies (CT) under Award Number 2302908. BCA, DGCA, PEV acknowledge the Yale Center for Research Computing for allocation of computation time to perform the dynamics simulations on the Grace cluster. DGCA acknowledges Alexander Soudackov for helpful discussions regarding the adenine-thymine potential.
\end{acknowledgement}

\begin{suppinfo}
Additional description of the Hamiltonian mapping, the simulation protocol, benchmarks with experimental observations and dynamics figures for different dissipation parameters and computational basis size.
\end{suppinfo}

\section{Code and Data Availability}
The Python code for the Hamiltonian analysis and dynamical simulations can be found at \href{https://github.com/dcabral00/cQED4ChemDyn}{https://github.com/dcabral00/cQED4ChemDyn}. 
The data files used to generate the figures in the manuscript are also provided at \href{https://github.com/dcabral00/cQED4ChemDyn}{https://github.com/dcabral00/cQED4ChemDyn} and hosted at \href{https://doi.org/10.5281/zenodo.13826722}{https://doi.org/10.5281/zenodo.13694461}. 

%-------------------------------------
% Bibliography
%-------------------------------------

\bibliography{references.bib}

\clearpage
% \appendix
\include{supporting_information}

\end{document}

%% file: supporting_information.tex
\begin{center}
\bf{
{\large Supporting Information for\\ 
"A Roadmap for Simulating Chemical Dynamics On A Parametrically Driven Bosonic Quantum Device"}}
\end{center}

\subsection{Mapping the  Device Hamiltonian to Chemical Double-Well}

We start out by considering the following general Hamiltonian which is suitable for modeling the dynamics of a wide range of elementary chemical reactions:
\begin{align}
    \hat{H}_{DW}    & = \frac{\hat{p}^2}{2m} +  k_4 \hat{x}^4 - k_2 
    \hat{x}^2+k_1 \hat{x}~~. \label{eq:dwps_SI}
\end{align}
Here, $\hat{x}$ and $\hat{p}$ are the position and momentum operators associated with motion along  the reaction coordinate, which satisfy $[\hat{x} , \hat{p} ] = i \hbar$;
$m$ is the mass associated with the motion along the reaction coordinate; and $\{ k_1, k_2, k_4 \}$ are positive and real parameters whose values define the double well free energy profile, and thereby the specific chemical system, that the Hamiltonian describes. More specifically, given the double-well free energy profile for a specific chemical system, which can be obtained from electronic structure and MD simulations,\cite{vazquez_entropic_2014}
we assume that it can be fitted to a minimal fourth-order polynomial of the form 
$V(x)=k_4 x^4 - k_2 x^2 + k_1 x$. The $k_4 x^4 - k_2 x^2$ term is necessary for obtaining the double-well feature, while the $k_1 x$ term is necessary in order to account for asymmetry between the reactant and product wells ($k_1=0$ gives rise to a symmetrical double-well free energy profile, which corresponds to an iso-energetic chemical reaction for which $\Delta G =0$). It should be noted that a third order $x^3$ term is excluded. This is necessary for mapping onto the Hamiltonian of currently accessible experimental Kerr-cat devices (see below), and justified by the fact that adding a $x^3$ term is not necessary for capturing the main features associated with a chemical reaction, namely an asymmetrical double-well profile. It should also be noted that a description of the chemical dynamics in terms of a TST/Arrhenius-like rate constant requires coupling the
reaction coordinate to a thermal bath of nonreactive DOF in order to make activation to the transition state and barrier crossing possible, followed by equilibration in the product well before significant recrossing can occur (see below).
\begin{figure}
    \centering
    \includegraphics[width=1.0\textwidth]{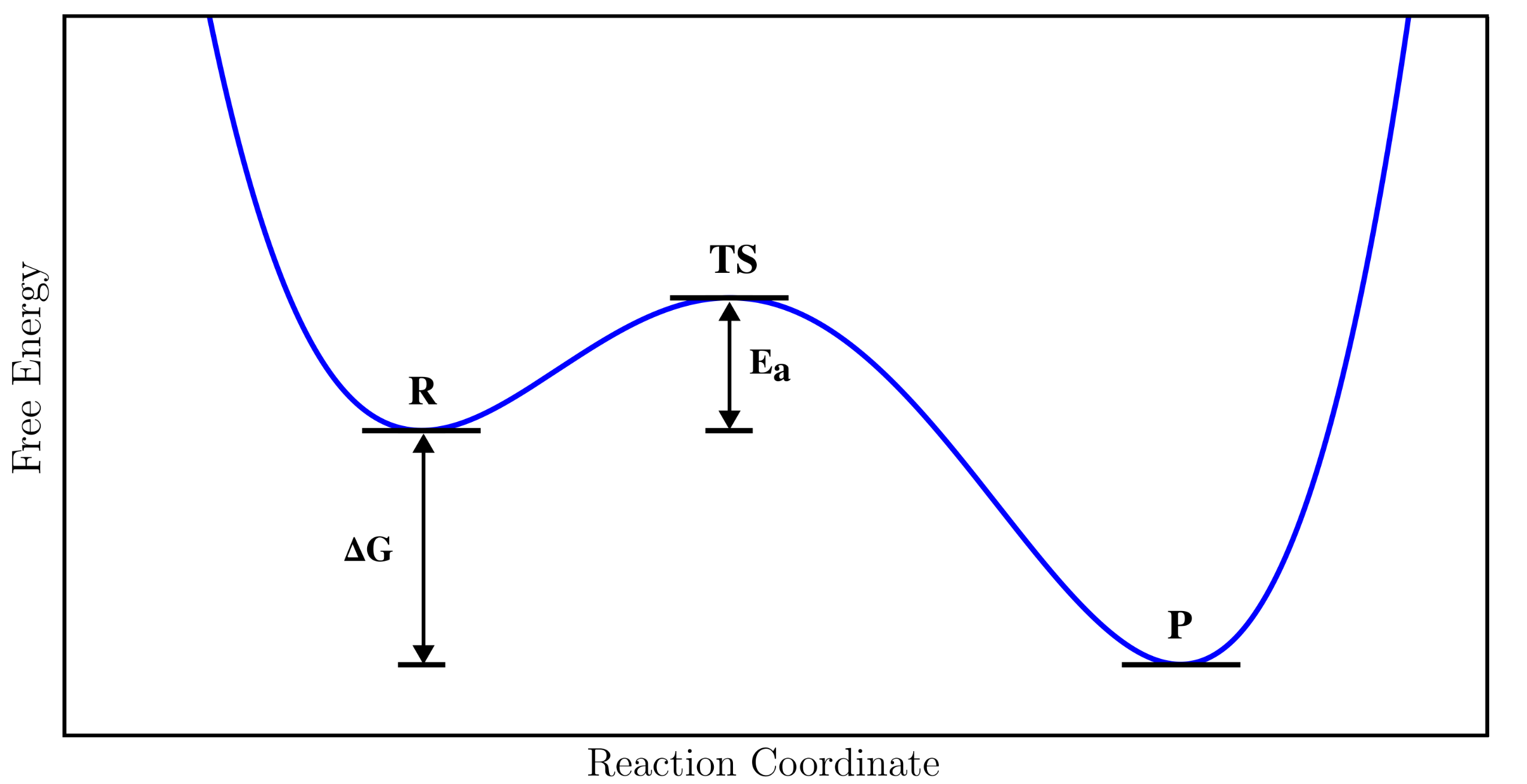}
    \caption{A schematic view of the the free energy double-well profile, $V(x)$, along the reaction coordinate, $x$. The reactant and product wells are designated by $R$ and $P$, respectively. The transition state, which corresponds to the barrier top, is designated by $TS$. $E_a$ and $\Delta G$ are the activation energy and  reaction free energy, respectively. It should be noted that the reaction coordinate needs to be coupled to a thermal bath of nonreactive DOF (not shown) in order for rate kinetics to be emerge.
}
    \label{fig:schematic}
\end{figure}
We consider the effective Hamiltonian for currently experimentally
realizable parametrically-driven Kerr-cat cQED devices, which is given
by:\cite{frattini_squeezed_2022, venkatraman_driven_2023, Reynoso2023,
chavez-carlos_spectral_2023, 
Iachello2023, Gonzalez2024}
\begin{equation}
    \hat{H}_{KC}
    = \Delta \hat{a}^{\dagger} \hat{a} - K 
    (\hat{a}^{\dagger})^2 (\hat{a})^2 + \epsilon _2 (\hat{a}^2 +
    \hat{a}^{\dagger 
    2}) + \epsilon _1 (\hat{a} + \hat{a}^{\dagger})~~.
    \label{eq:kerr-cat_SI}
\end{equation}
Here, $\hat{a}$ and $\hat{a}^\dagger$ are (unit-less) photonic creation and
annihilation operators associated with the electromagnetic mode supported by
the cavity, which satisfy $[\hat{a},\hat{a}^\dagger]=\hat{1}$, and 
$\{ \Delta,K,\epsilon _2,\epsilon _1 \}$ are experimentally controllable
parameters (all given in terms of energy units). 
Noting that the double-well and Kerr-cat Hamiltonians in Eqs. (\ref{eq:dwps_SI})
and (\ref{eq:kerr-cat_SI}), respectively, are both given by fourth-order
polynomials determined by four free parameters ($\{ m, k_1 , k_2, k_4 \}$ and
$\{ \Delta,K,\epsilon _2,\epsilon _1 \}$, respectively),
our goal in the next step is to map the Kerr-cat Hamiltonian in Eq.
(\ref{eq:kerr-cat_SI}), onto the chemical double-well Hamiltonian in Eq.
(\ref{eq:dwps_SI}). 

To this end, we first need to map the photonic operators, $\{ \hat{a} ,
\hat{a}^\dagger\}$ onto the operators associated with motion along the reaction
coordinate, $\{ \hat{x} , \hat{p} \}$. To generate the correct dynamics, the
mapping needs to be consistent with the corresponding commutators: 
$[\hat{x} , \hat{p} ] = i \hbar$ and $[\hat{a} , \hat{a}^\dagger ] = \hat{1}$.
A mapping that satisfies this is given by:
\begin{align}
    \hat{a} &=\frac{1}{\sqrt{2}}\left( 
    \frac{1}{c}\hat{x}+\frac{ic}{\hbar}\hat{p}\right) 
    ~~~;~~~
    \hat{a}^\dagger =\frac{1}{\sqrt{2}}\left( 
    \frac{1}{c}\hat{x}-\frac{ic}{\hbar}\hat{p}\right)
    \nonumber
    \\
    \hat{x} &= \frac{c}{\sqrt{2}} \left( \hat{a} + \hat{a}^\dagger \right)
    ~~~;~~~ 
    \hat{p} = \frac{\hbar}{i \sqrt{2} c} 
     \left( \hat{a} - \hat{a}^\dagger \right)~~.
     \label{eq:aaxp_mapping_SI}
\end{align}
Here, $c$ is a constant parameter that has units of length (same units as
$\hat{x}$).
Importantly, the value of $c$ is arbitrary in the sense that the commutators
$[\hat{x} , \hat{p} ] = i \hbar$ and $[\hat{a} , \hat{a}^\dagger ] = \hat{1}$
are invariant to the choice of $c$. In other words, the mapping of $\{ \hat{a}
, \hat{a}^\dagger\}$ onto $\{ \hat{x} , \hat{p} \}$ is not unique. As we will
see below, this flexibility with respect to the choice of $c$ plays an crucial
role in mapping the Kerr-cat Hamiltonian in Eq. (\ref{eq:kerr-cat_SI}), onto the
chemical double-well Hamiltonian in Eq. (\ref{eq:dwps_SI}). 

Substituting the expressions for $\hat{a}$ and $\hat{a}^\dagger$ in terms of
$\hat{x}$ and $\hat{p}$ from Eq. (\ref{eq:aaxp_mapping_SI}) into Eq.
(\ref{eq:kerr-cat_SI}), we can recast the {\em negative} of the Kerr-cat
Hamiltonian in terms of the $\hat{x}$ and $\hat{p}$ (dropping constant terms
which do not impact the dynamics):
\begin{align}
    -\hat{H}_{KC} &= \frac{c^2}{\hbar^2} \left( \epsilon_2 - K 
    -\Delta/2\right)\hat{p}^2 
    +\frac{K}{4c^4} \hat{x}^4 
    - \frac{1}{c^2}\left( 
\epsilon_2+K+\Delta/2\right)\hat{x}^2 
    -\frac{\epsilon_1 
    \sqrt{2}}{c}\hat{x}
    \notag 
    \label{eq:kcps_SI}\\ &+ \frac{Kc^4}{4\hbar^4}\hat{p}^4 
    +\frac{K}{4\hbar^2}\hat{x}^2\hat{p}^2+\frac{K}{4\hbar^2}\hat{p}^2\hat{x}^2
\end{align}
Comparing Eq. (\ref{eq:kcps_SI}) with Eq. (\ref{eq:dwps_SI}), we see that while
$-\hat{H}_{KC}$ contains $\hat{p}^2$, $\hat{x}^4$, $\hat{x}^2$ and $\hat{x}$
terms which can be mapped onto the corresponding terms in the chemical
double-well Hamiltonian in Eq. (\ref{eq:dwps_SI}), it also contains spurious
$\hat{p}^4$, $\hat{x}^2\hat{p}^2$ and $\hat{p}^2\hat{x}^2$ terms that lack
counterparts in Eq. (\ref{eq:dwps_SI}). 

In the next step, we map  
$\{ \Delta,K,\epsilon _2,\epsilon _1 \}$ onto $\{ m, k_1 , k_2, k_4 \}$ by
requiring consistency between the $\hat{p}^2$, $\hat{x}^4$, $\hat{x}^2$ and
$\hat{x}$ terms in Eqs. (\ref{eq:dwps_SI}) and (\ref{eq:kcps_SI}), which leads to
the following mapping relations:
\begin{align}
    K &=  4c^4 k_4 \label{eq:first1} 
    \\
    \epsilon_2 &= \frac{\hbar^2 + 
    \label{K_eq}
    2c^4 k_2 m}{4c^2 m} \\
    \Delta &=\frac{2c^4 k_2 m-\hbar^2 -   
    16c^6 k_4 m}{2c^2 m}\\
    \epsilon_1 &= -\frac{c k_1}{\sqrt{2}} \label{eq:last}
\end{align}
We also take advantage of the aforementioned flexibility in choosing the value
of $c$ to minimize the effect of the spurious $\hat{p}^4$, $\hat{x}^2\hat{p}^2$
and $\hat{p}^2\hat{x}^2$ terms in Eq. (\ref{eq:dwps_SI}). As we show below, doing
so requires that we choose a value of $c$ small enough so that it satisfies the
following inequality:
\begin{align}
\frac{\hbar^2}{ m k_2 c^4} \gg 1
\label{eq:ineq1_SI}
\end{align}

To derive the inequality in Eq. (\ref{eq:ineq1_SI}), we note that the $\hat{x}^4$,
$\hat{x}^2$ and $\hat{x}$ terms in Eq. 
(\ref{eq:kcps_SI}) become {\em larger} relative to the other terms with {\em
decreasing} $c$. This suggests that choosing a sufficiently small value of $c$
can make the spurious $p^4$, $x^2 p^2$ and $x^2 p^2$ terms negligible. However,
the fact the kinetic energy term in Eq. (\ref{eq:kcps_SI}), $\frac{c^2}{\hbar^2}
\left( \epsilon_2 - K 
    -\Delta/2\right)\hat{p}^2 $, 
also decreases with decreasing $c$ implies that the value of $c$ also needs to
be chosen such that the spurious terms will be negligible compared to it. It
must be noted that if one puts Eq. \ref{eq:first1}-\ref{eq:last} in units of
$K$, that being $\{\epsilon_1 /K, \epsilon_2 /K, \Delta/K \}$, these quantities
diverge when $\lim c\rightarrow 0$, with these quantities getting quite large
when $c$ is small. So, it is necessary to pick a value of $c$ which produces
experimentally accessible values of $\{\epsilon_1 /K, \epsilon_2 /K, \Delta/K
\}$ for a given chemical system while ensuring sufficient chemical accuracy.

To this end, we consider the the symmetrical double-well case 
($k_1=0$), for which it can be shown that the reactant and product equilibrium
geometries are given by $\pm x_0$, where
$x_0 = \sqrt{\frac{k_2}{2 k_4}}$
and the activation energy is given by
$E_a = \frac{k_2^2}{4 k_4}$.
Thus, $\{ x_0,E_a \}$ are interchangeable with $\{ k_2,k_4 \}$ in this case,
such that Eq. (\ref{eq:first1})
becomes $K=\frac{4 E_a c^4}{x_0^4}$. Hence,
\begin{equation}
\frac{K}{4\hbar^2}\left[ \frac{c^4}{\hbar^2}\hat{p}^4 
+\hat{x}^2\hat{p}^2+\hat{p}^2\hat{x}^2 \right] 
\rightarrow
\frac{E_a c^4}{\hbar^2 x_0^4}\left[ \frac{c^4}{\hbar^2}\hat{p}^4 
+\hat{x}^2\hat{p}^2+\hat{p}^2\hat{x}^2
\right]
\end{equation}
Given that $x_0$ set the length scale of the chemical system, one can estimate
the order of magnitude of the $\hat{x}^2 \hat{p}^2$ and $\hat{p}^2 \hat{x}^2$
terms to be 
$\frac{E_a c^4}{\hbar^2 x_0^4} x_0^2\hat{p}^2 = 
\frac{E_a c^4}{\hbar^2 x_0^2} \hat{p}^2$.
Thus, requiring that the spurious $\hat{x}^2 \hat{p}^2$ and $\hat{p}^2
\hat{x}^2$ terms are negligible relative to the kinetic energy term,
$\frac{\hat{p}^2}{2m}$
gives rise to the inequality
$\frac{E_a c^4}{\hbar^2 x_0^2} \hat{p}^2 \ll \frac{1}{2m} \hat{p}^2$, which can
be rearranged to give 
$\frac{\hbar^2 x_0^2}{2m E_a c^4} \gg 1$.
Noting that $k_2=\frac{2E_a}{x_0^2}$ then leads to the inequality in Eq.
(\ref{eq:ineq1_SI}). 

The fact that the $\hat{p}^4$ term scales like $c^4$, while the $\hat{p}^2$
term scales like $c^2$, also implies that the $\hat{p}^4$ will become
negligible for a sufficiently small value of $c$. 
In fact, the same inequality, Eq. (\ref{eq:ineq1_SI}), can be derived by noting
that 
$\frac{1}{2m}\hat{p}^2 \gg \frac{E_a c^8}{\hbar^4 x_0^4} \hat{p}^4$ is
equivalent to $\frac{1}{2m} \gg \frac{E_a c^8}{\hbar^4 x_0^4} \hat{p}^2$
and that 
the momentum is maximal when the particle is around the minima, where the
potential energy can be approximated as being harmonic. Invoking the virial
theorem for the harmonic  oscillator,
according to which the
expectation values of the kinetic energy is equal to that of the potential
energy, and noting that 
$E_a$ sets the potential energy scale for the chemical system, we can then
estimate $\hat{p}^2$ by $2m E_a$ in the inequality $\frac{1}{2m} \gg \frac{E_a
c^8}{\hbar^4 x_0^4} \hat{p}^2$, which turns it into the inequality 
$\left( \frac{\hbar^2 x_0^2}{2m E_a c^4} \right)^2 \gg 1$. Thus, satisfying the
inequality in Eq. (\ref{eq:ineq1_SI}), which is equivalent to $\frac{\hbar^2
x_0^2}{2m E_a c^4} \gg 1$,
also guarantees that the $\hat{p}^4$ term will becomes negligible compared the
the $\hat{p}^2$ kinetic energy term. 

Finally, the same argument would also hold for an {\em
asymmetrical} double-well since the length and
energy scales of the chemical system are not going to be significantly affected
by the addition of the asymmetry. 

%%%%%%%%%%%%%%%%%%%%%%%%%%%%%%%%%%%%%%%%%%%%%%%%%%%%%%%%%%%%%%%%%%%%%
%% The manuscript does not need to include \maketitle, which is
%% executed automatically.
%%%%%%%%%%%%%%%%%%%%%%%%%%%%%%%%%%%%%%%%%%%%%%%%%%%%%%%%%%%%%%%%%%%%%
\subsection{Computational Methods}
\label{sec:Methods}
In this work, we examine the dissipative dynamics of the asymmetric Kerr-cat
Hamiltonian,
\begin{equation}
\label{eq:H_kc}
   \dfrac{\hat{H}}{\hbar} = \Delta \hat{a}^{\dagger} \hat{a} - K
   (\hat{a}^{\dagger})^2 (\hat{a})^2 + \epsilon _2 (\hat{a}^2 +
   \hat{a}^{\dagger 2}) + \epsilon _1 (\hat{a} + \hat{a}^{\dagger}) ,
\end{equation}
where $\Delta$, $\epsilon _2$, $\epsilon _1$ control the potential landscape
parameters such as inter-well separation, barrier height, and well asymmetry,
respectively. The entire Hamiltonian is scaled by $K$, which is taken to be a
constant value throughout the manuscript, unless otherwise stated. The
operators $a^{\dagger}, a$ are the quantum Harmonic oscillator excitation and
de-excitation operators expressed in the basis of Fock states.
To simulate the dissipative dynamics of this Hamiltonian, we use the Lindblad
master equation:
\begin{equation}
   \dfrac{\partial \hr (t)}{\partial t} = - \dfrac{i}{\hbar} \left[ \hat{H},
   \hr(t) \right] + 
    \mathcal{D}  \left[ \hr (t) \right] ,
\end{equation}
where $\hat{H}$ is the Kerr-cat Hamiltonian, $\hr(t)$ is the time-dependent
density matrix and 
$\mathcal{D}[\hr (t)]$ is the dissipator defined as:
\begin{equation}
    \mathcal{D}[\hr (t)] = \kappa(1+n_{\text{th}}) \left(\hat{a} \hr
    \hat{a}^{\dagger} - \dfrac{1}{2}\{\hat{a}^{\dagger}\hat{a}, \hr\}
    \right) + \kappa n_{\text{th}} \left(\hat{a}^{\dagger} \hr \hat{a} -
    \dfrac{1}{2}\{\hat{a} \hat{a}^{\dagger}, \hr\} \right)
\end{equation}
with $a^{\dagger}, a$ being excitation and deexcitation operators, whose
effect is governed by the magnitude of the thermal parameters $\kappa$ and
$n_{\text{th}}$.
To implement the Lindblad equation and simulate dissipative dynamics, we
vectorize the density matrix and matricize the Lindbladian, using the
relationship $\text{vec}(AXB) = (B^T \otimes A) \text{vec}(X)$, such that 
\begin{equation}\label{eq:differential_Lindbladian}
   \dhr = \lind \hr
\end{equation}
Thus, we must find a suitable representation for $\lind$. We introduce
identity matrices to utilize the vectorization relationship and apply it to
the Hamiltonian:
\begin{align}\label{eq:hamiltonian_matricized}
   \left[ \hat{H}, \hr(t) \right] &= \hat{H} \hr \mathbb{I} - \mathbb{I}
   \hr \hat{H}\\
       &= \left( \mathbb{I} \otimes \hat{H} - \hat{H}^{T} \otimes \mathbb{I}
       \right) \hr
\end{align}
Similarly, we can alter the dissipator:
\begin{align}\label{eq:dissipator_matricized}
   \mathcal{D}[\hr (t)] = \kappa(1+n_{\text{th}}) &\left(\hat{a} \hr
   \hat{a}^{\dagger} 
   - \dfrac{1}{2}\left( \hat{a}^{\dagger}\hat{a} \hr \mathbb{I} +
   \mathbb{I} \hr \hat{a}^{\dagger}\hat{a} \right) \right) \nonumber \\
   + \kappa n_{\text{th}} &\left(\hat{a}^{\dagger} \hr \hat{a}
   -\dfrac{1}{2}\left( \hat{a}\hat{a}^{\dagger} \hr \mathbb{I} + \mathbb{I}
   \hr \hat{a} \hat{a}^{\dagger} \right) \right) \\
   =  \kappa(1+n_{\text{th}}) &\left(\hat{a}^{*} \otimes \hat{a}
   - \dfrac{1}{2}\left( \mathbb{I} \otimes \hat{a}^{\dagger}\hat{a} +
   \hat{a}^{T} \hat{a}^{*} \otimes \mathbb{I} \right) \right) \hr \nonumber
   \\
   + \kappa n_{\text{th}} &\left(\hat{a}^{T} \otimes \hat{a}^{\dagger}
   -\dfrac{1}{2}\left(  \mathbb{I} \otimes \hat{a}\hat{a}^{\dagger} +
   \hat{a}^{*} \hat{a}^T \otimes \mathbb{I}  \right) \right) \hr
\end{align}
Reassembling the complete matricized Lindblad equation, we obtain
\begin{align}\label{eq:lindbladian_matricized}
   \dot{\hr} &= \Bigg( \mathbb{I} \otimes \hat{H} - \hat{H}^{T} \otimes
   \mathbb{I} \nonumber \\
   &+ \kappa(1+n_{\text{th}}) \left(\hat{a}^{*} \otimes \hat{a}
   - \dfrac{1}{2}\left( \mathbb{I} \otimes \hat{a}^{\dagger}\hat{a} +
   \hat{a}^{T} \hat{a}^{*} \otimes \mathbb{I} \right) \right) \nonumber \\
   &+ \kappa n_{\text{th}} \left(\hat{a}^{T} \otimes \hat{a}^{\dagger}
   -\dfrac{1}{2}\left(  \mathbb{I} \otimes \hat{a}\hat{a}^{\dagger} +
   \hat{a}^{*} \hat{a}^T \otimes \mathbb{I}  \right) \right) \Bigg) \hr
\end{align}
To simulate the population dynamics, we integrate equation
\ref{eq:differential_Lindbladian},
\begin{align}
   \hr _t = \hr (t) &= e^{\lind t} \hr _0
\end{align}
and calculate the action of time-evolution of the Lindbladian operator on the
propagated density matrix for a small time-step $\tau=0.1$:
\begin{equation}
   \hr (t) = e^{\lind \tau} \hr (t-\tau)
\end{equation}
The matrix exponential operator is implemented using the
\verb|scipy.linalg.expm| routine, which implements a scaling and squaring
algorithm based on Pade's approximation.
\cite{scipy_expm_exponential_propagator}
%%%%%%%%%%%%%%%%%%%%%%%%%%%%%%%%%
\subsection{Initial State}
\label{sec:initial_state}
\begin{figure}[ht]
    \centering
    \includegraphics[width=1.0\textwidth, clip, trim = 0 0 0 
    0]{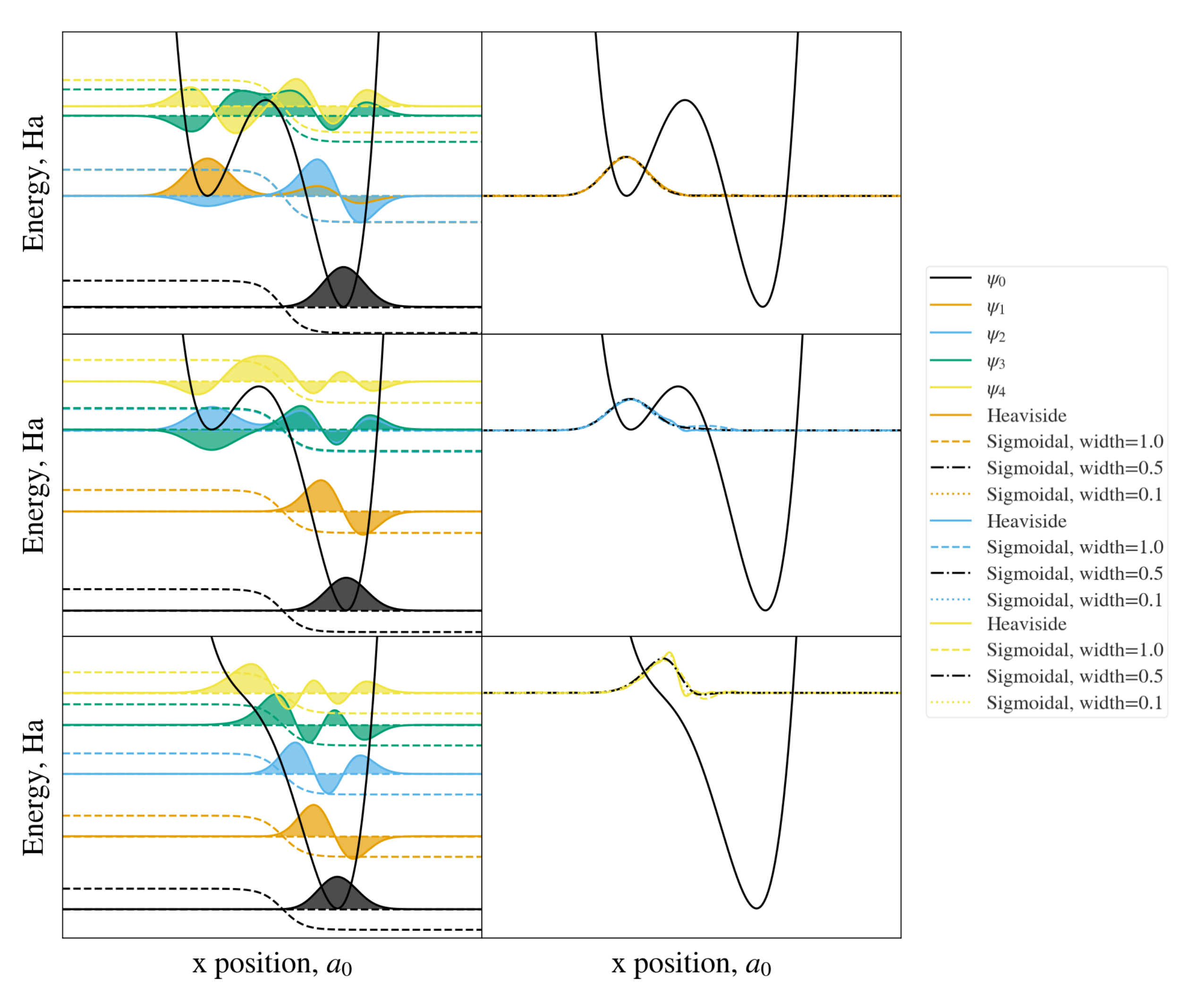} 
    \caption{Initial state selection for three different values of $\epsilon
    _1$, $\epsilon _2$ by applying a sigmoidal or Heaviside filtering function.
    The procedure is illustrated for the first 5 eigenstates, plotted with the
    metapotential on the background to showcase how the localization scheme
    performs.
    The right side showcases the effect of the different filter parameters as
    applied to the most suitable state that contains more than 50\% density on
    the top well.
    Higher values of the sigmoidal tail value reduce initial state
    localization, while higher values introduce oscillatory motions due to the
    verticality of the filter function near its center.}
    \label{fig:init_state}
\end{figure}

The initial state of the system is crucial to the dynamics. We start by
diagonalizing the Hamiltonian $H$ to find the eigenstate matrix $\Phi$ of the
system in the harmonic oscillator Fock basis:
\begin{equation}
    H\Phi = \lambda \Phi
\end{equation}
We then find the grid-based position representation ($x$) of the individual
eigenstates, $\phi_i$, using the quantum Harmonic oscillator basis set:
\begin{equation}
    \phi_i = \sum _n ^N c_{n,i} \psi _n (x)
\end{equation}
Where $c_{n,i}$ indicate the expansion coefficients associated with eigenstate
$i$ and using $N$ harmonic oscillator functions of the form
\begin{equation}
    \psi _{n}(x)={\frac {1}{\sqrt {2^{n}\,n!}}}\left({\frac {m\omega }{\pi
    \hbar }}\right)^{1/4}e^{-{\frac {m\omega x^{2}}{2\hbar }}}H_{n}\left({\sqrt
    {\frac {m\omega }{\hbar }}}x\right),\qquad n=0,1,2,\ldots
\end{equation}
In this expression, $n$ indicates the order of the basis function, $m$
represents the mass, $\omega$ is the fundamental frequency of the oscillator,
$\hbar$ is the reduced Planck's constant and $H_n$ are the physicist's Hermite
polynomials of order $n$, which follow the following recurrence relation:
\begin{equation}
    \begin{cases}
        H_0 (x) = 1 \\
        H_1 (x) = 2x \\
        \vdots \\
        H_{n+1} (x) = 2x H_n (x) - 2nH_{n-1} (x)
    \end{cases}\nonumber
\end{equation}
We select a suitable initial state by finding the first state with more than
50\% amplitude on the desired portion of the potential energy surface (figure
\ref{fig:init_state}, left panel) and then convolve it with a sigmoidal filter
function of the form:
\begin{equation}
    S (x; x_0, t) = \dfrac{1}{1+e^{-(x-x_0)/t}},
\end{equation}
% ------------------------------
where $x$ indicates the position, $x_0$ indicates the cutoff position and $t$
the smoothness of the function near the cutoff.
We observe that the Heaviside function, $\Theta (x; x_0) = 1$ if $x \geq x_0$ else $\Theta (x; x_0) = 0$ , is recovered when
taking $\lim _{t \rightarrow 0} S (x; x_0, t)$.
This allows localization of the initial state in position space which we then
convert back to the Fock basis representation (figure \ref{fig:init_state},
right panel).
All dynamics trajectories use eigenstate selection with a sigmoidal filter with
a tail of 0.5.
\clearpage
\subsection{Dynamics Subspace}
The accuracy of the dynamics is dependent on the number of Fock basis states
used. However, the size of the Lindbladian matrix scales as $\mathcal{O}(N^4)$
with the number of Fock states as compared to the Hamiltonian
$\mathcal{O}(N^2)$, which makes matrix exponentiation (performed once for each
set of Hamiltonian parameters) and multiplication (performed for each timestep
in a trajectory) a limiting factor in simulation. Thus, we generate the
complete Hamiltonian with a large number of Fock states (N=300) and numerically
diagonalize to obtain accurate eigenvalues and eigenvectors and use the first
M=20 states to perform a similarity transformation matrix to reduce the
dynamics computational space:
\begin{equation}
    H _{N \times N} C _{N \times N} = \lambda _{N \times N} C _{N \times N}
    \rightarrow D _{N \times M} \equiv C _{N \times M}
\end{equation}
Where $\lambda _{N \times N}$, $C _{N \times N}$ contains the eigenvalues,
eigenvectors of the Hamiltonian $H _{N \times N}$ in the full $N$-dimensional
space and $C _{N \times M}$ represent the reduced dimensionality eigenvector
matrix containing the first $M$ eigenstates which is defined as the
transformation matrix $D _{N \times M}$. Then the initial state ($\hr _0$),
the Hamiltonian ($H$) and the Lindbladian ladder operators ($a^{\dagger}, a$)
are transformed into the reduced Hilbert space according to the transformation, 
 \begin{equation}
     A'_{M \times M} = D ^T_{M \times N} A _{N \times N} D _{N \times M}
 \end{equation}
 As a consequence of this, the ladder operators now encode information about
 the properties of the Hamiltonian and thus can better simulate the dynamics of
 the system.
\subsection{Observables}\label{sec:observables}
For this work, we focus on observables corresponding to traces with the
time-evolved state. These include traces with the initial state corresponding
to the lowest-lying state on the initial well, $\text{Tr}\{{\hr_t\hr_0}\}$, and
traces with the Heaviside function to obtain the population on the right
side, $P_R=\text{Tr}\{\hr_t \Theta (x; x_0) \}$ or traces with the complement of the
Heaviside function to obtain the population on the left, $P_L = \text{Tr}\{\hr_t (1-\Theta (x; x_0))\}$.
Finally, we look at the eigenvalues obtained by exact diagonalization of the
Lindbladian to assess the principal modes/mechanisms of population transfer as
well as the long-time final equilibrium state. We focus on the maximum amplitude non-zero
real eigenvalue, to compute the decay time defined as follows:
\begin{equation}
T_X = -[\Re{\lambda}]^{-1}
\end{equation}
which represents the slower decaying timescale of the Lindbladian. Note that this gives qualitative
insight into the relaxation rate, while bypassing the more expensive 
requirement of performing dynamics propagation.

\clearpage
\subsection{Simulating barrier crossing dynamics on a Kerr-cat device}

Barrier crossing dynamics of the type typically observed in chemical systems
requires coupling the reaction coordinate to a bath of non-reactive DOFs which
acts both as an energy source for activating the chemical system from the
bottom of the reactant well to the vicinity of the barrier top and as an energy
sink for stabilizing the system in the product well once barrier crossing
occurred. Since the Kerr-cat Hamiltonian in Eq. (\ref{eq:kerr-cat_SI}) only
describes the dynamics along the reaction coordinate, treating it as a closed
quantum system undergoing unitary dynamics would not generate the desirable
chemical dynamics. Coupling the reaction coordinate to a thermal bath of
nonreactive DOFs takes us to the domain of {\em nonunitary} open quantum
systems dynamics. 
In what follows, we will assume that this dynamics is described by the
following Lindblad quantum master equation:
\begin{align}
    \dfrac{\partial \hr (t)}{\partial t} &= - \dfrac{i}{\hbar} \lrb{\hkc, \hr (t)} \\
        &+ \kappa \lrp{1 + \nth} \lrb{\ha \hr (t) \hadag 
            - \dfrac{1}{2} \lrB{\hadag \ha,\hr (t)} } \\
        &+ \kappa \nth \lrb{\hadag \hr (t) \ha
            - \dfrac{1}{2} \lrB{\ha\hadag,\hr (t)} } \nnn
            &\equiv \mathcal{L} \hr(t). \\
        \label{eq:lindblad}
\end{align}

Here, $\hat{\rho}(t)$ is the density operator that describes the state of the
reactive system, $\hat{H}_{KC}$ is the Kerr-cat Hamiltonian of the reactive
system [Eq. (\ref{eq:kerr-cat_SI})], 
$\{ \kappa n_{\text{th}} , \kappa (n_{\text{th}} +1)\}$ are parameters that
determine the rates of bath-induced uphill and downhill transitions,
respectively, and thereby the coupling strength between the reaction coordinate
and the bath of non-reactive DOFs and $\mathcal{L}$ is the Lindbladian
superoperator. 

Simulating the dissipative dynamics described by Eq. (\ref{eq:lindblad}) 
was accomplished by vectorizing the density operator and matricizing the
Lindbladian superoperator, followed by diagonalizing $\mathcal{L}$ and
propagating the vectorized density operator according to
\begin{equation}
 \hat{\rho} (t) = e^{\mathcal{L} t} \hat{\rho} (0)~~. 
\end{equation}

The initial state was chosen so that it is localized in the reactant well. 
To this end, we picked the first eigenstate of the Kerr-cat Hamiltonian with
more than 50\% probablity of being in the reactant well and then multiplied it
by a sigmoidal  function that filtered out the part of the wave function that
resides in the product well. Given this reactant-well-localized wave function,
$| \psi_R \rangle$, the initial density operator is given by $\hat{\rho} (0) =
| \psi_R \rangle \langle \psi_R |$ (a pure state). 

\begin{figure}[ht!]
    \centering
    \includegraphics[width=1.0\textwidth, clip
    ]{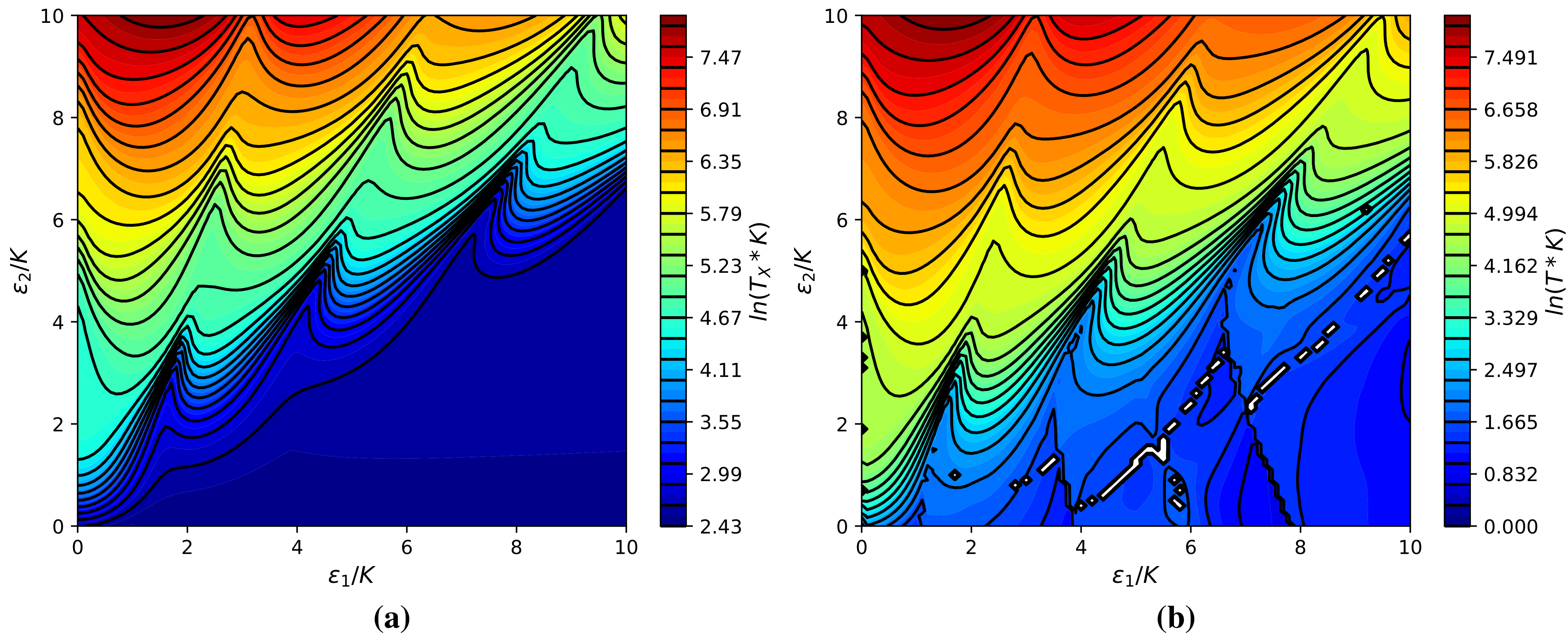}
    \caption{Relaxation timescales associated with the (a) Lindbladian maximal
    real
    eigenvalue (left) and (b) the dynamical relaxation rates (right) obtained
    by fitting the population traces as a function 
    of the Kerr-cat parameters $\epsilon_1$ and $\epsilon_2$.
    Both plots use dissipation parameters $\kappa=0.1$, $n_{\text{th}}=0.1$. }
    \label{fig:summary_Tx_regimes}
\end{figure}

The barrier crossing rate constant is given by $k = 1/T_X$, where $T_X$ defines
the barrier crossing time scale. $T_X$ was determined in two ways:
\begin{enumerate}
\item As the inverse of the maximum amplitude non-zero
real eigenvalue of the
    Lindbladian supermatrix, obtained via diagonalization,
        as in figure \ref{fig:summary_Tx_regimes}a.  
\item As the time scale of decay of $Tr \lrb{\hr _t \Theta_X}$, obtained via fitting to an exponential, as in figure
        \ref{fig:summary_Tx_regimes} (b). 
\end{enumerate}
The two methods for determining $T_X$ gave similar results and were found to exhibit
the same 
behavior when it comes to the dependence of $T_X$ on the Kerr-cat parameters.
The results 
reported in the text were obtained via method 1 unless otherwise noted (figure 
\ref{fig:summary_Tx_regimes}b). 

The following analysis is for the case of $\Delta=0$, and using the convention
$\hbar=1$ and $K$ as a unit of energy.
A complete description of the methodology is included in 
\nameref{sec:Methods}.

%%%%%%%%%%%%%%%%%%%%%%%%%%%%%%%%%%%%%%%%%%%%%%%%%%%%%%%%
\clearpage
%%%%%%%%%%%%%%%%%%%%%%%%%%%%%%%%%%%%%%%%%%%%%%%%%%%%%%%%

The dependence of $T_X$ on 
the Kerr-cat parameters $\epsilon_1$ and $\epsilon_2$ is shown in 
Fig. \ref{fig:summary_Tx_regimes}.
The plot shows a rich structure including (a) a zone in the lower right corner
where the barrier crossing is very fast, which corresponds to a low barrier or
a complete lack of a barrier, 
(b) Fast barrier crossing in the upper left corner for  
particular values of ($\epsilon_1,\epsilon_2$)
where the energy levels in the reactant and product wells are in resonance
(see white lines in Figs. 
\ref{fig:summary_Tx_regimes} and SI), and (c) alternation 
between ``broad" and ``thin" resonance transitions both as a function of 
$\epsilon_1$ for fixed $\epsilon_2$ and along the  
($\epsilon_1,\epsilon_2$) resonance line.

%%%%%%%%%%%%%%%%%%%%%%%%%%%%%%%%%%%%%%%%%%%%%%%%%%%%%%%%
\clearpage
\begin{figure}[ht!]
    \centering
    \includegraphics[width=1.0\textwidth, clip, trim = 0 0 0 
    0]{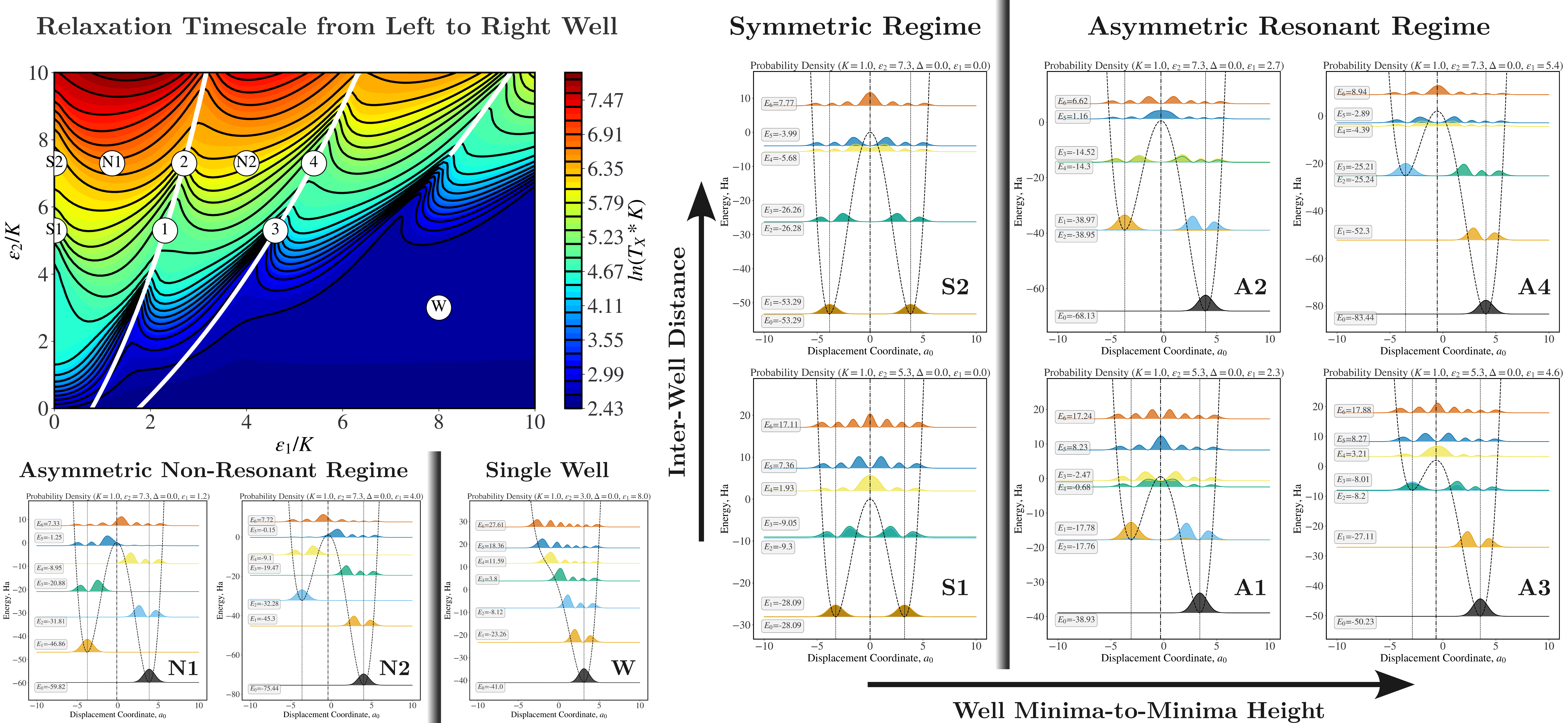}
    \caption{Resonant and non-resonant regimes of the eigendensities between 
    both sides of the double-well along the $p=0$ metapotential cut. Each state 
    are plotted as a function of position superimposed on the metapotential cut
    spanning $x\in [-10,10]$ Bohr. The 
    vertical axis denotes the absolute energy obtained by diagonalizing the 
    Hamiltonian. Panels include the asymmetric non-resonant regime (N1, N2), a
    region where the double-well description is no longer valid (W). The
    resonant regime falls along the white lines and reflect regimes with no
    asymmetry (S1, S2) as well as regions of increasing well asymmetry (A1-A2,
    A3-A4) by changing the minima-to-minima height.}
    \label{fig:summary_Tx_regimes_densities}
\end{figure}
%%%%%%%%%%%%%%%%%%%%%%%%%%%%%%%%%%%%%%%%%%%%%%%%%%%%%%%%
The fast 
barrier crossing regimes correspond to resonances between the energy levels in
the reactant and product wells 
which lead to efficient 
tunneling through the barrier.
The aforementioned ``thin" regions
correspond to eigenstate overlap 
near the top of the barrier as well as a high state density at the barrier
top, 
providing a transient state to retain population before decaying to the global
ground 
state (see figure \ref{fig:summary_Tx_regimes} and corresponding panels S1, 2,
3 of 
figure \ref{fig:summary_Tx_regimes_densities}). By contrast the ``broad" regimes 
contain degenerate states but no density centered at the top of the barrier,
thus 
reducing the overall population transfer rate (see figure 
\ref{fig:summary_Tx_regimes} and corresponding panels S2, 1, 4 of figure 
\ref{fig:summary_Tx_regimes_densities}). However, the rate is nonetheless 
enhanced due to the presence of quasi-degenerate states near the barrier top, 
providing a pathway for barrier crossing. Furthermore, the intermediate states
without resonant states have longer lifetimes due to lack of overlap between 
eigenstates or localized state density on either side of the well (see figure 
\ref{fig:summary_Tx_regimes} and corresponding panels N1 and N2 of figure 
\ref{fig:summary_Tx_regimes_densities}). Going along each of the resonance 
lines showcases an increasing number of degenerate pairs of states with 
increasing $\epsilon _2$ (figure \ref{fig:summary_Tx_regimes}: S1-S2; 1-2; 
3-4). Going between different resonance lines with increasing $\epsilon _1$ 
changes the first state index in resonance between the two wells (S1: first and 
second, 1: second and third, 3: third and fourth). 

Finally, a decrease in the minimum-to-minimum height decreases the barrier 
height until the double-well is destroyed (figure 
\ref{fig:summary_Tx_regimes_densities}, W). This decreases the relaxation 
lifetime as the kinetics driving the process is merely vertical de-excitation
to the ground state. Although there can be a relaxation timescale associated
with 
the process this is not a measurement of the kinetics of population transfer
between the wells, as this is an ill-defined process in this parameter regime.

\begin{figure}[ht!]
    \centering
    \includegraphics[width=1.0\textwidth,
    clip]{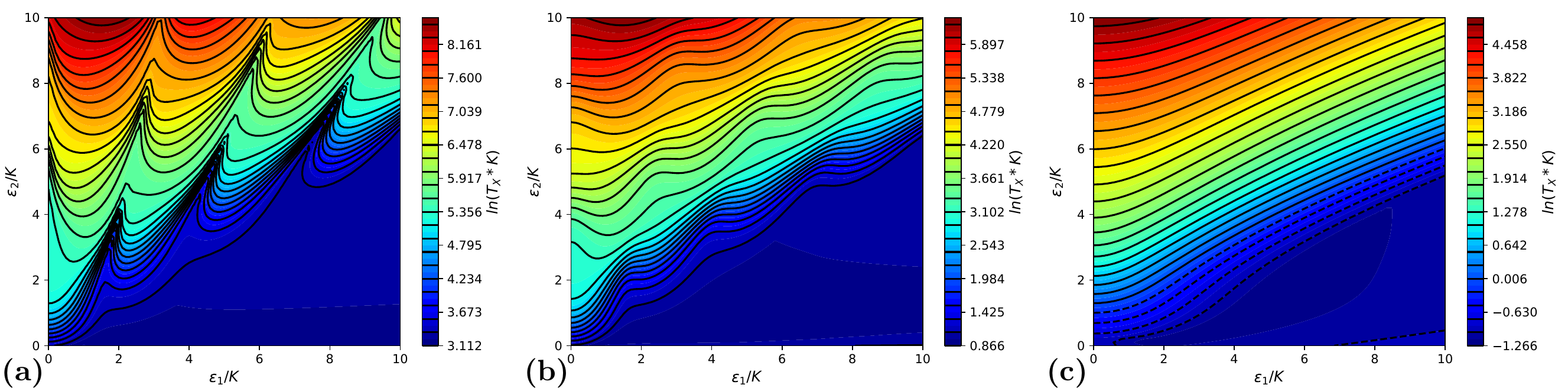}
    \caption{Dependence of timescale obtained by exact Lindbladian 
    diagonalization on the dissipation parameter $\kappa$, for 
    $n_{\text{th}}=0.1$.
    (a) $\kappa=0.05$.
    (b) $\kappa=0.5$.
    (c) $\kappa=5.0$.
    }
    \label{fig:Lindbladian_kappa}
\end{figure}

Fig. \ref{fig:Lindbladian_kappa} shows how the dependence of $T_X$ on
$\epsilon_1$ and $\epsilon_2$ is impacted by the
the strength of coupling between the reaction coordinate and the bath of
nonreactive DOFs, as measured by $\kappa$. 
As expected, $T_X$ shows an overall trend of increasing with decreasing
coupling strength, which can be traced back to the fact that the rate of
activation from the bottom of the reactant well to to the vicinity of the
barrier top and stabilization in the product well after barrier crossing are
determined by $\kappa$. 
Additionally, the dependence of $T_X$ on $\epsilon_1$ and $\epsilon_2$ is seen
to become less structured with increasing $\kappa$, which can be traced back to
ability of dissipation to wash out resonance effects.  
More specifically, while
tunnelling dominates the kinetics at low values of $\kappa$,
classical-like barrier crossing and thereby TST/Arrhenius-like kinetics is
observed at larger values of $\kappa$. 
A more extensive set of data
that showcases this observation for a wider range of parameter regimes is
provided in the supporting information.

\begin{figure}[ht!]
    \centering
    \includegraphics[width=1.0\textwidth,
    clip]{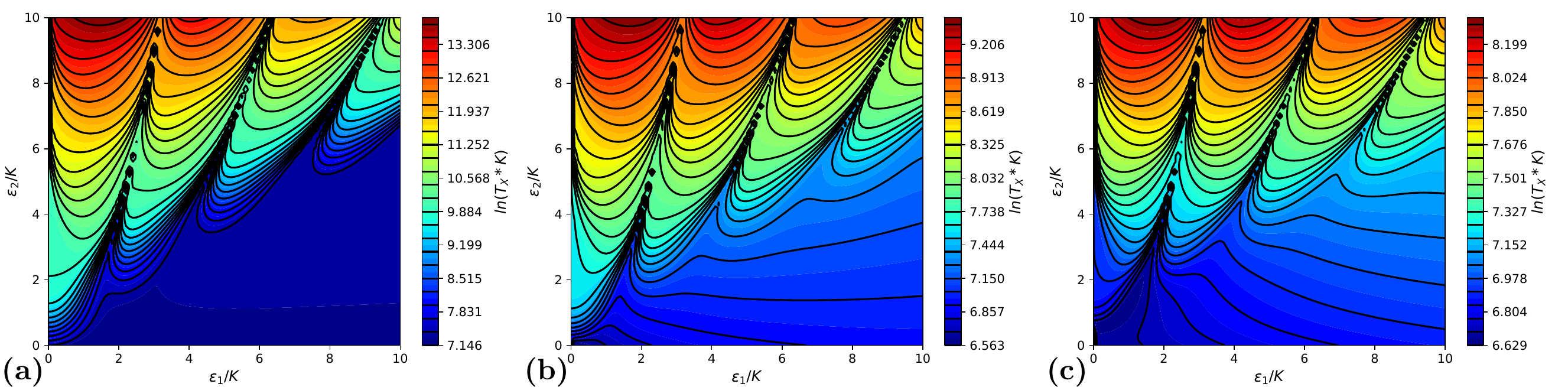}
    \caption{Dependence of timescale obtained by exact Lindbladian 
    diagonalization on the dissipation parameter $\kappa$, for $\kappa=0.001$.
    (a) $n_{\text{th}}=0.05$.
    (b) $n_{\text{th}}=0.5$.
    (c) $n_{\text{th}}=1.0$.
    }
    \label{fig:Lindbladian_kappa0.001_nbar}
\end{figure}

Finally, Fig. \ref{fig:Lindbladian_kappa0.001_nbar} shows how the dependence of
$T_X$ on $\epsilon_1$ and $\epsilon_2$ is impacted by the bath temperature, as
measured by $n_{\text{th}}$. 
While $T_X$ shows an overall increase with decreasing 
$n_{\text{th}}$, the structure is seen to be minimally impacted by changing $n_{\text{th}}$. 

% ------------------------------
\subsection{Basis Set Convergence}

\begin{figure}[ht]
  \centering
  \includegraphics[width=.45\textwidth, clip, trim = 10 0 20
    20]{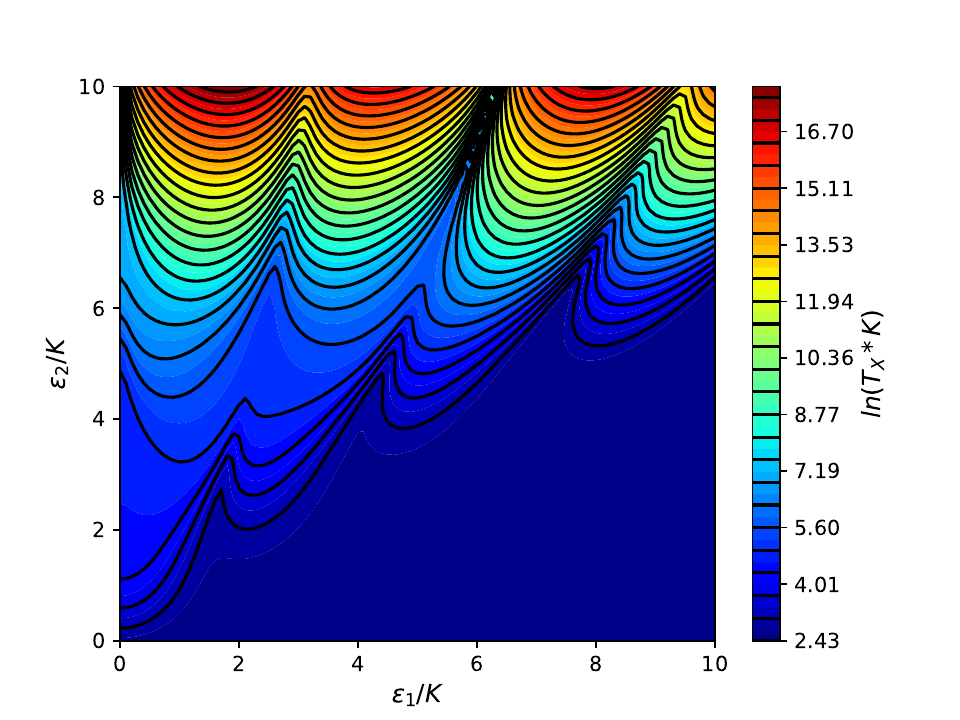}
  \includegraphics[width=.45\textwidth, clip, trim = 10 0 20
    20]{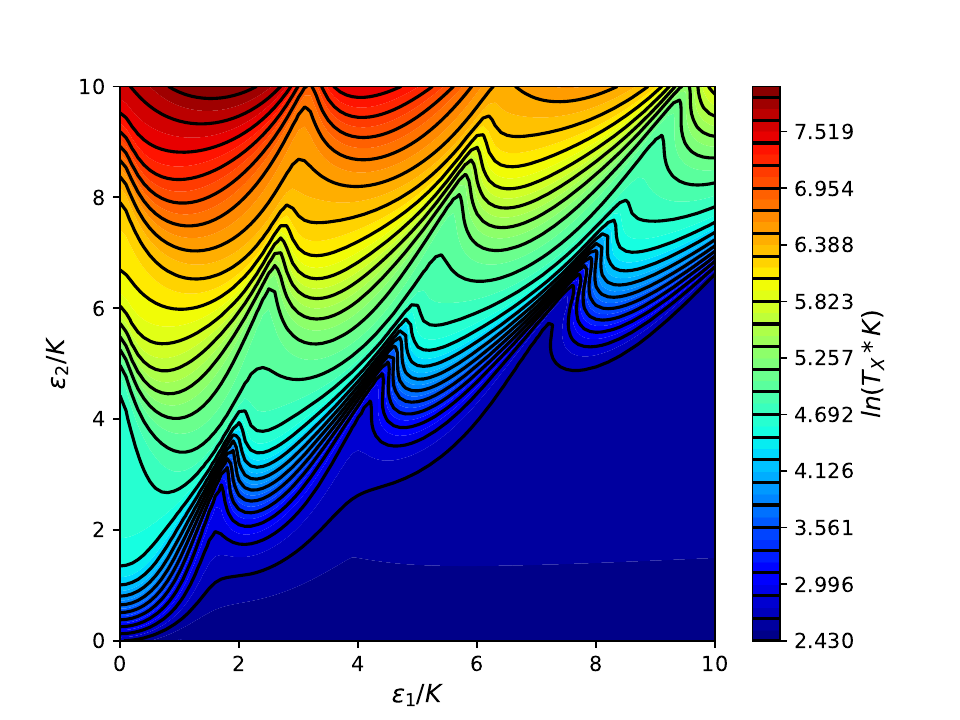} 
  \includegraphics[width=.45\textwidth, clip, trim = 10 0 20
    20]{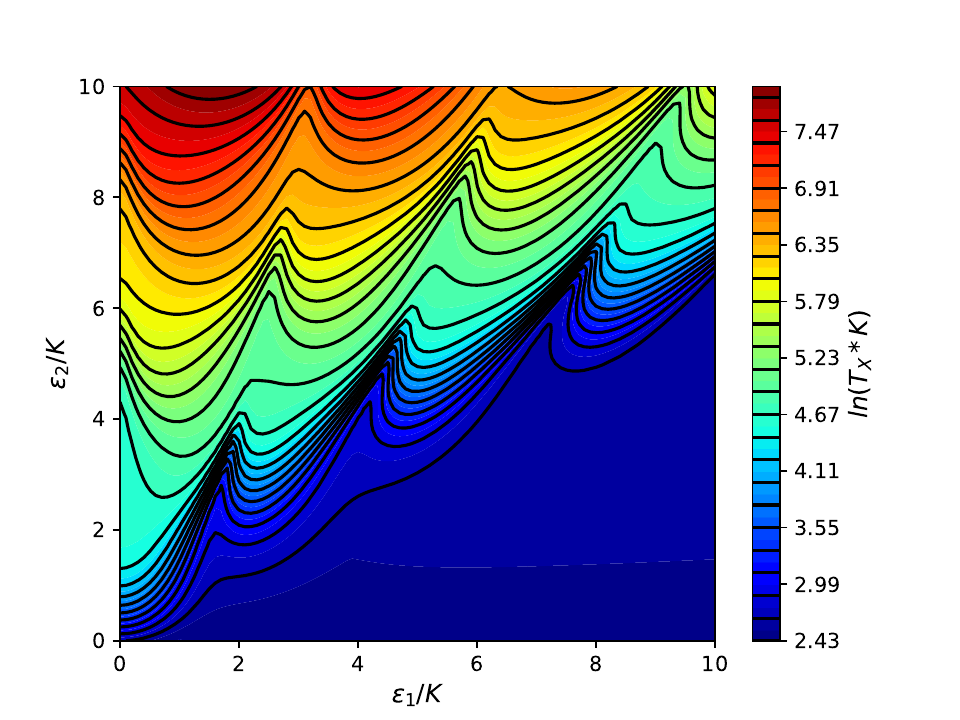} 
  \includegraphics[width=.45\textwidth, clip, trim = 10 0 20
    20]{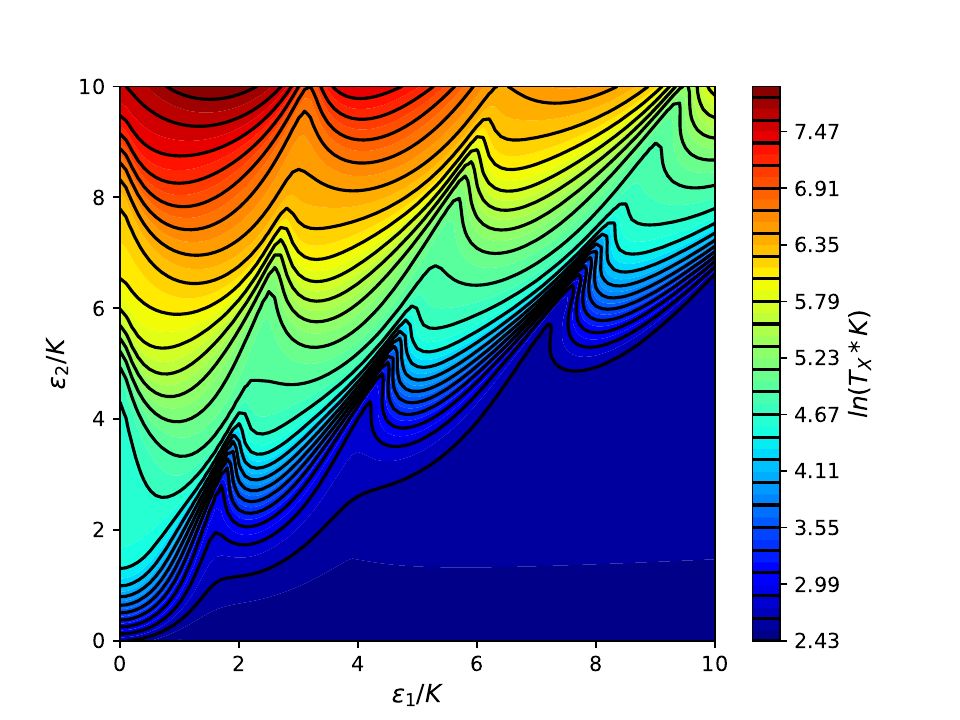}
    \caption{Convergence of Lindbladian eigenvalue timescale as a function of
    the number of Eigen-basis. (a) $n_{basis}=5$. (b) $n_{basis}=10$. (c)
    $n_{basis}=20$. (d) $n_{basis}=30$.  }
  \label{fig:Lindbladian_convergence}
\end{figure}

In this section, we explore the convergence of the Kerr-cat Hamiltonian parameters to
the number of Eigen-basis used. 
In Fig. \ref{fig:Lindbladian_convergence}, we present the timescales obtained
from the exact diagonalization of the Lindbladian as a function of Hamiltonian
parameters $\epsilon_1$ and $\epsilon_2$, for a different number of
Eigen-basis.
As can be appreciated, with $n_{basis}>10$ the timescales are semi-quantitative
converged, and with $n_{basis}>20$ quantitative agreement is found. Unless
otherwise stated, we used $n_{basis}=20$ for the device dynamics.

\clearpage
\subsection{Chemical Potentials}

The chemical potentials were obtained from the literature, and are listed within
the tables in this section.
\begin{table}[]
    \centering
    \begin{tabular}{|c|c|c|c|c|c|} \hline System & $k_4$ & $k_3$ & $k_2$ & 
        $k_1$ & Units \\ \hline 
        Thymine-Adenine (DNA)\cite{AT_potential} & 0.02068986 & 0.00525515 &  0.0413797 & 0.0157655 & $E_h$ \\ 
        Malonaldehyde (cis-trans)\cite{ghosh_dynamics_2012,ghosh_optimised_2015} & 
        0.00009374 & 0.000109 & 0.00299 & 0.005232 & a.u. \\ 
        Malonaldehyde (cis-cis)\cite{ghosh_dynamics_2012,ghosh_optimised_2015} & 0.000714286 & 0 & 0.004 & 0 & a.u. \\ \hline \end{tabular}
    \caption{Literature chemical potential parameters given by expression 
    $V_{\text{literature}} = k_1 x - k_2 x^2 - k_3 x^3 + k_4 x^4$}
    \label{tab:lit_params1}
\end{table}

\begin{table}[]
    \centering
    \begin{tabular}{|c|c|c|c|c|c|c|c|} \hline System & $V_1$ & $V_2$ & $a_1$ & 
        $a_2$ & $r_1$ & $r_2$ & Units \\ \hline Guanine-Cytosine (DNA)
        \cite{slocombe_open_2022} & 
        0.1617 & 0.082 & 0.305 & 0.755 & -2.7 & 2.1 & a.u. \\ \hline \end{tabular}
        \caption{Literature chemical potential parameters given by expression 
        $V_{\text{literature}} = V_1\{ 
        \exp(-2a_1[x-r_1])-2\exp[-a_1(x-r_1)]\}+V_2\{ 
        \exp(2a_2[x-r_2])-2\exp[a_2(x-r_2)]\}$}
        \label{tab:lit_params2}
\end{table}
Furthermore, we note that the adenine-thymine potential is expressed in terms
of a unitless length parameter $\zeta = x/x_0$, which has been estimated to
be $x_0=1.9592 \, a_0$ based on matching the energy eigenvalues listed
in reference \citenum{AT_potential}.
For the dynamics shown in this work, the given literature potentials where fit 
to the simpler double-well potential, without a cubic polynomial term,
\begin{equation}
    \hat{V}_{\text{DW}} = k_4 \hat{x}^4 + k_2 \hat{x}^2 + k_1 \hat{x}_1
\end{equation}

\subsection{Dynamics Trajectories and Rate Fits}

This section lists additional trajectory plots for both the double-well and
Kerr-cat analog for different values of c and a plot of the resulting fitted
rates. These plots cover the dissipation parameters listed in the main text.

\begin{figure}[ht!]
    \centering
    \makebox[\textwidth][c]{
        \includegraphics[width=0.95\textwidth]{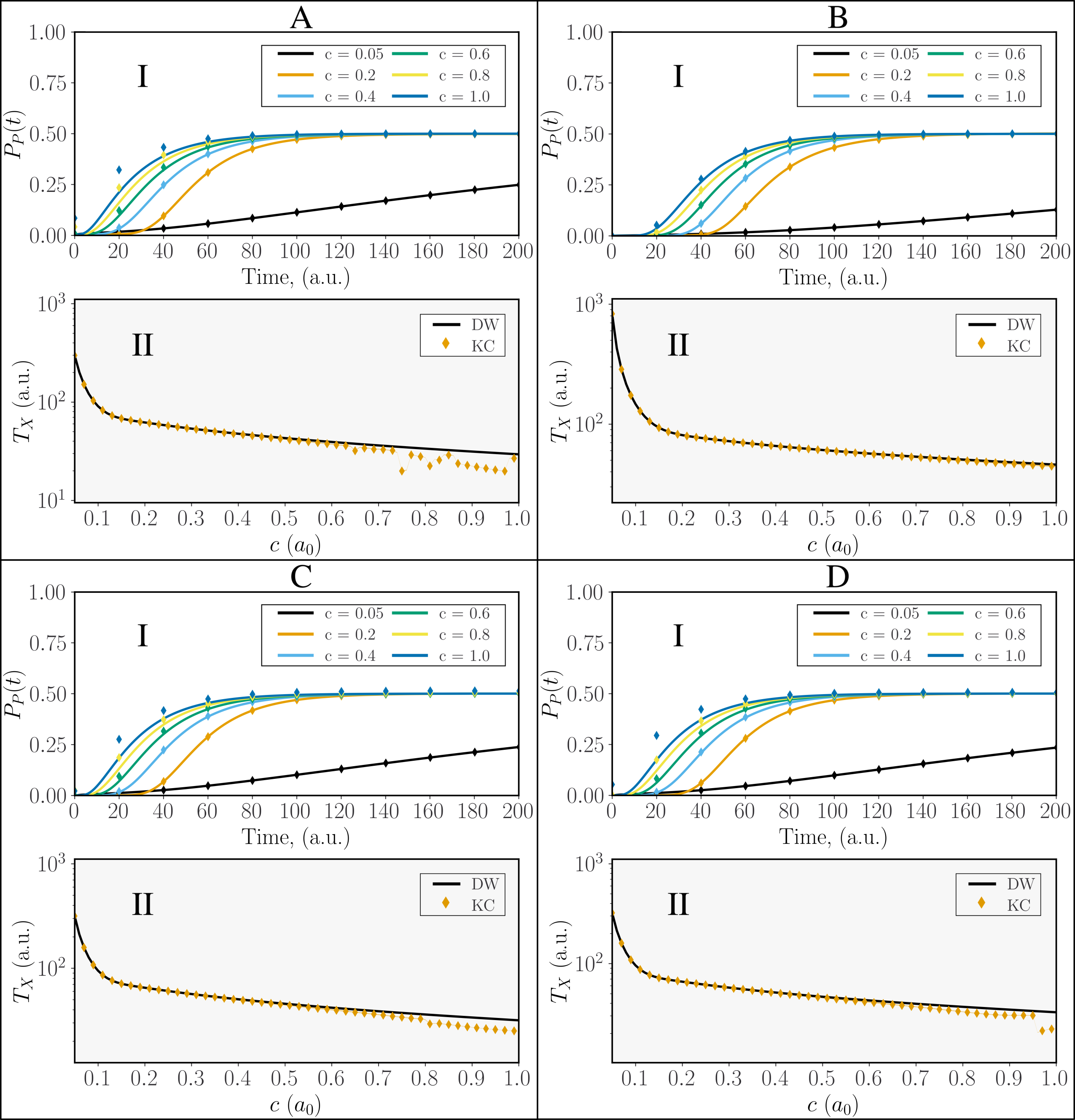}
    }
    \caption{Comparison of observables computed obtained with $\hat{H}_{DW}$ (solid lines) and $\hat{H}_{KC}$ (diamonds) as a function of $c$, using $\kappa=0.1$, and $\nth=0.1$. The time evolution of the product population for the four proton transfer reactions are shown in the top subpanels. The corresponding inverse reaction rate constants are shown in the bottom subpanels.
    }
    \label{fig:lindbladian_dynamics_k0.1_n0.1}
\end{figure}
\clearpage
\begin{figure}[ht!]
    \centering
    \makebox[\textwidth][c]{
        \includegraphics[width=0.95\textwidth]{
        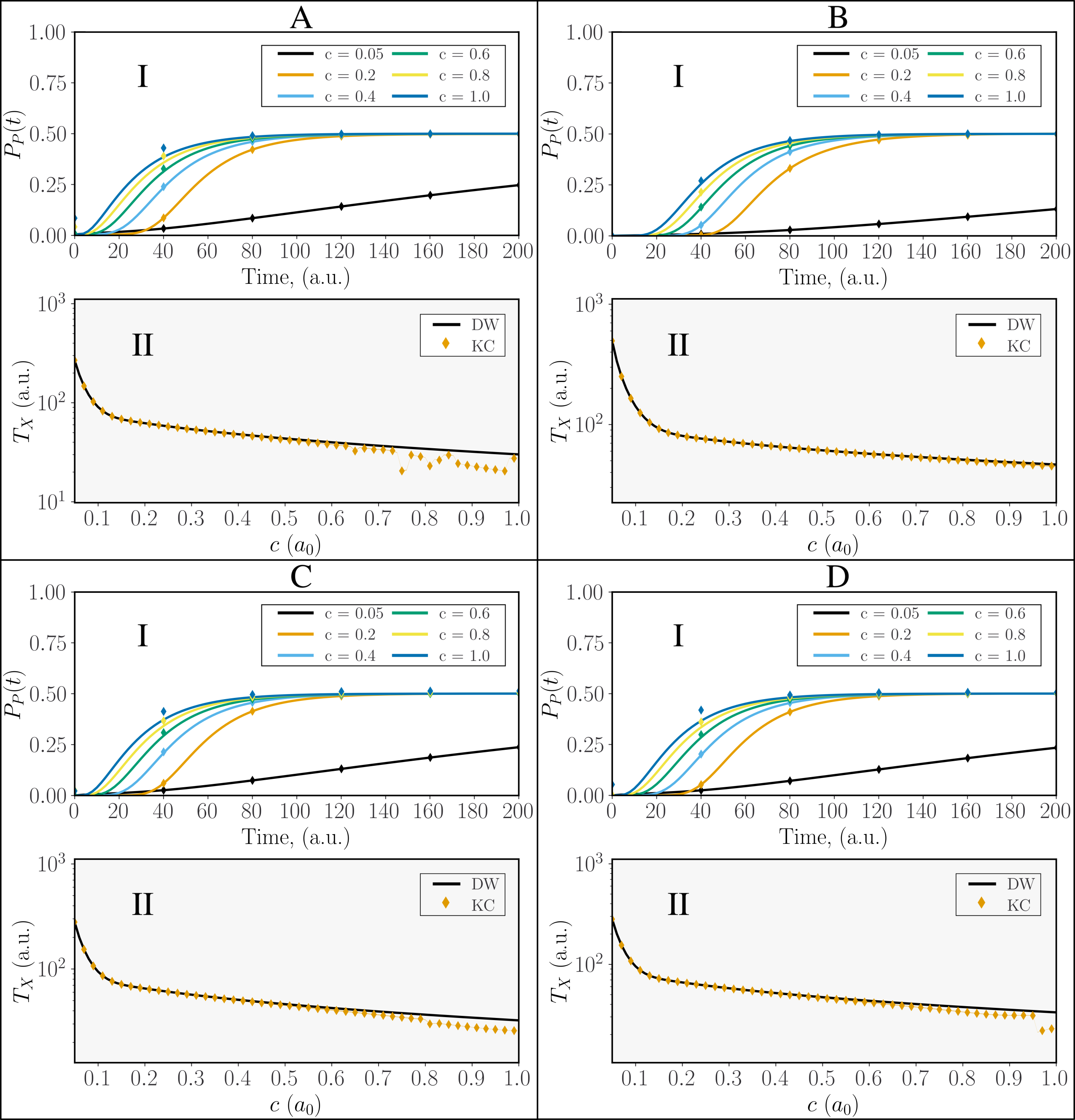}
    }
    \caption{Comparison of observables computed obtained with $\hat{H}_{DW}$ (solid lines) and $\hat{H}_{KC}$ (diamonds) as a function of $c$, using $\kappa=0.1$, and $\nth=0.05$. The time evolution of the product population for the four proton transfer reactions are shown in the top subpanels. The corresponding inverse reaction rate constants are shown in the bottom subpanels.
    }
    \label{fig:lindbladian_dynamics_k0.1_n0.05}
\end{figure}
\clearpage
\begin{figure}[ht!]
    \centering
    \makebox[\textwidth][c]{
        \includegraphics[width=0.95\textwidth]{
        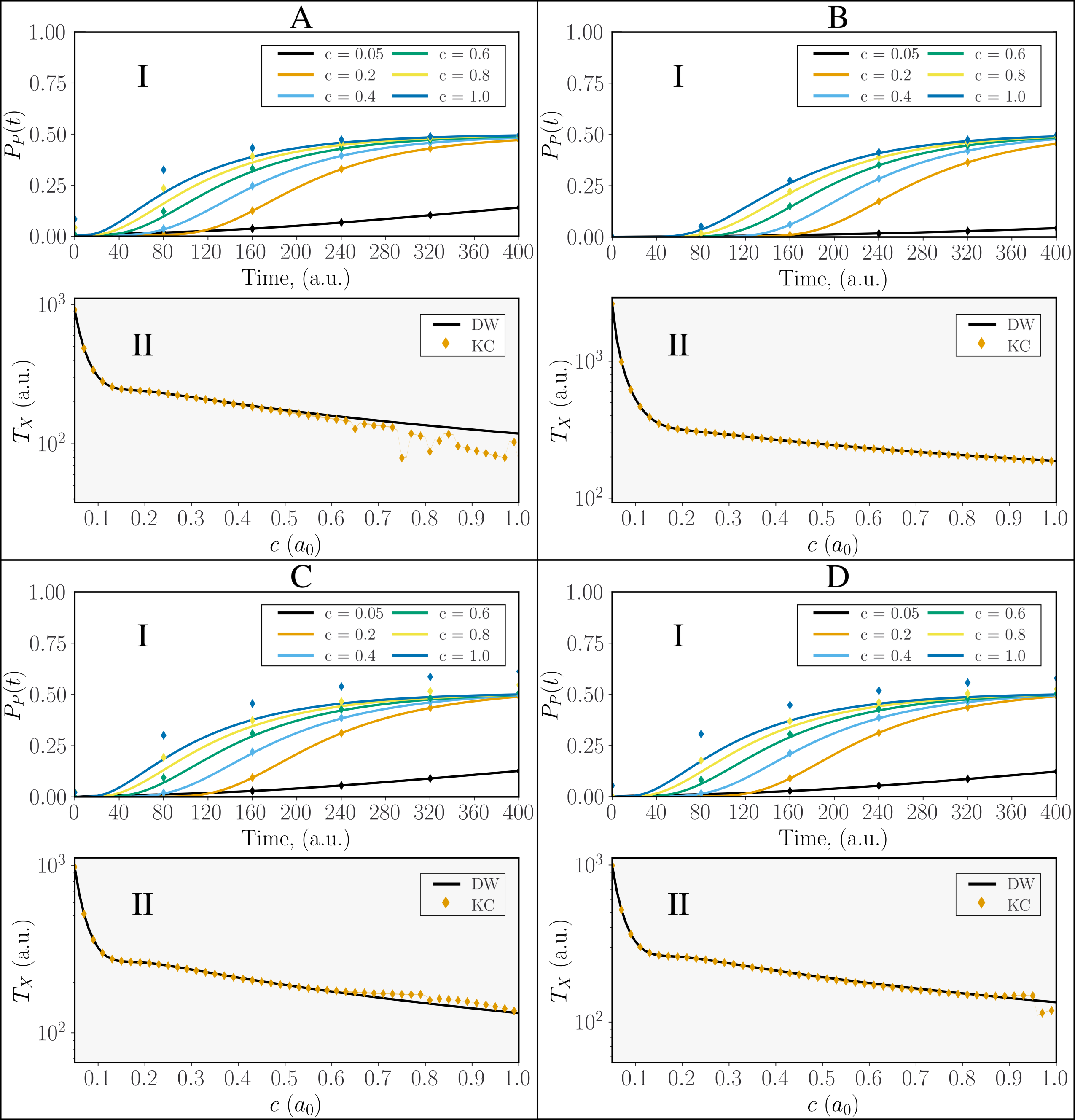}
    }
    \caption{Comparison of observables computed obtained with $\hat{H}_{DW}$ (solid lines) and $\hat{H}_{KC}$ (diamonds) as a function of $c$, using $\kappa=0.025$, and $\nth=0.1$. The time evolution of the product population for the four proton transfer reactions are shown in the top subpanels. The corresponding inverse reaction rate constants are shown in the bottom subpanels.
    }
    \label{fig:lindbladian_dynamics_k0.025_n0.1}
\end{figure}
\clearpage
\begin{figure}[ht!]
    \centering
    \makebox[\textwidth][c]{
        \includegraphics[width=0.95\textwidth]{
        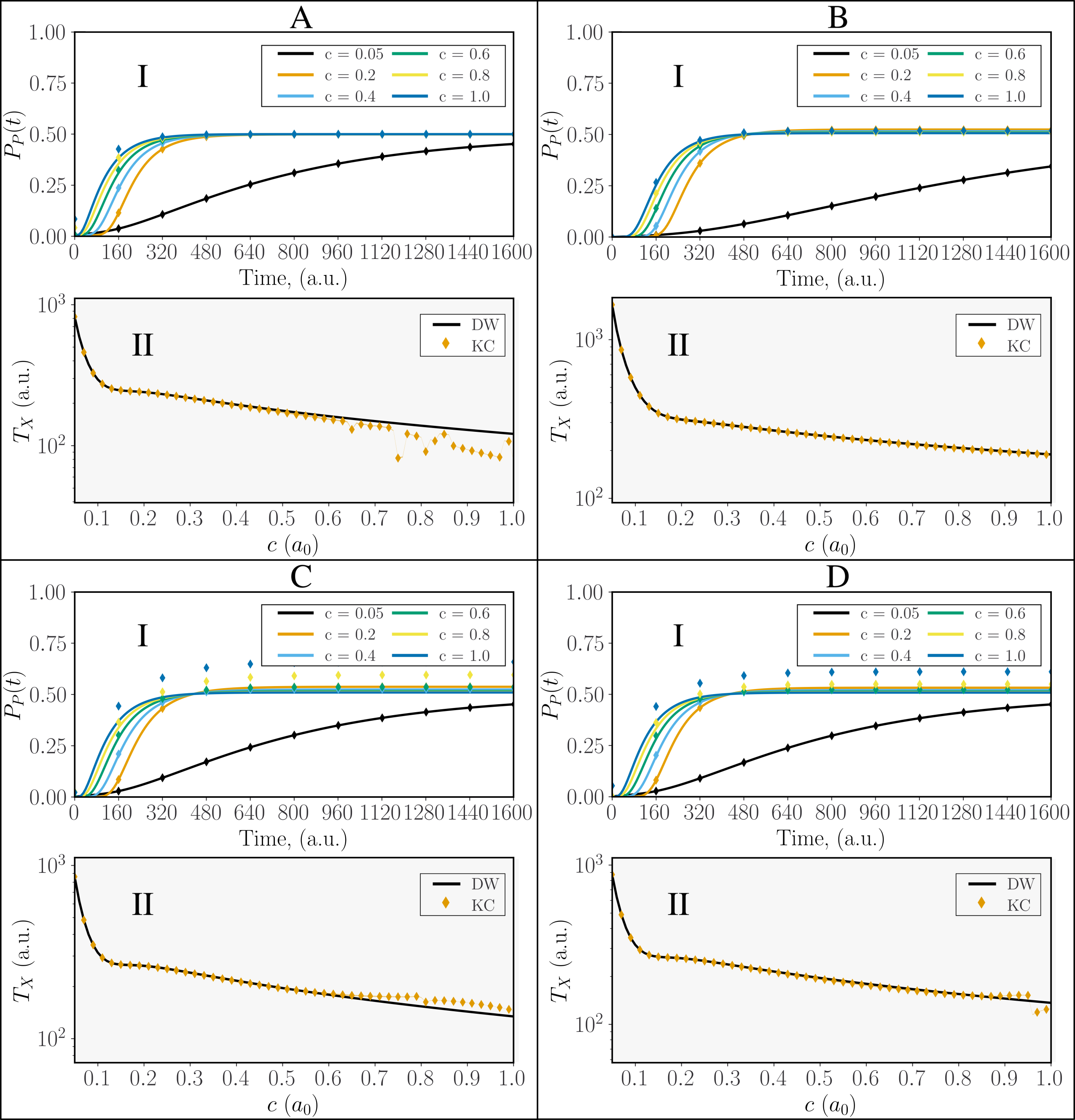}
    }
    \caption{Comparison of observables computed obtained with $\hat{H}_{DW}$ (solid lines) and $\hat{H}_{KC}$ (diamonds) as a function of $c$, using $\kappa=0.025$, and $\nth=0.05$. The time evolution of the product population for the four proton transfer reactions are shown in the top subpanels. The corresponding inverse reaction rate constants are shown in the bottom subpanels.
    }
    \label{fig:lindbladian_dynamics_k0.025_n0.05}
\end{figure}
\clearpage

\subsection{Basis Benchmark for Dynamical Evolution}
This section showcases the dynamics evolution figures for the four systems, using a different number of basis set to showcase the convergence of the results listed in the main text. The figures showcase the results of using a Fock space of dimension 50, 100 and 150 to accurately represent the Hamiltonian and eigenstates, which are then used for the dynamics propagation in a reduced subspace. For all cases, 50 eigenstates were used for the dynamics propagation, which incorporate many levels beyond the barrier top energy. For some of the listed systems, dynamics convergence with respect to number of eigenstates is observed for values much smaller than fifty. Deviations of the Kerr-cat fitted rates at larger $c$-values result from the Hamiltonian perturbative terms that are sensitive to the value of $c$ and errors associated with the fitting protocol.
\begin{figure}[ht!]
    \centering
    \makebox[\textwidth][c]{
        \includegraphics[width=0.95\textwidth]{
        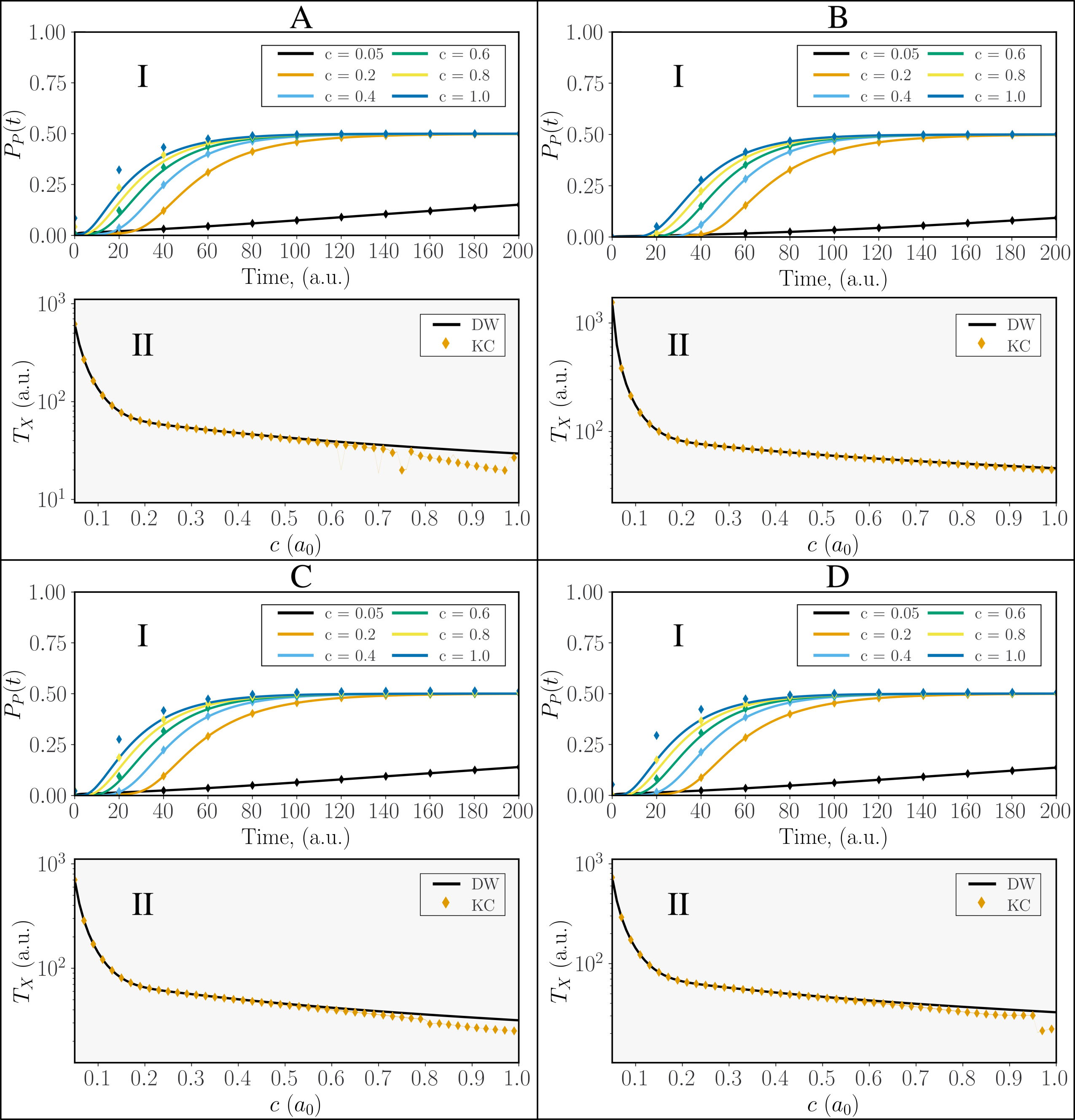}
    }
    \caption{Comparison of observables computed obtained with $\hat{H}_{DW}$ (solid lines) and $\hat{H}_{KC}$ (diamonds) as a function of $c$, using $\kappa=0.1$, and $\nth=0.1$. The time evolution of the product population for the four proton transfer reactions are shown in the top subpanels. The corresponding inverse reaction rate constants are shown in the bottom subpanels. These results were generated using 50 Fock basis for Hamiltonian and initial state preparation and a subspace of 50 eigenfunctions for dynamics propagation.
    }
    \label{fig:lindbladian_dynamics_k0.1_n0.1_NbF_50}
\end{figure}
\clearpage
\begin{figure}[ht!]
    \centering
    \makebox[\textwidth][c]{
        \includegraphics[width=0.95\textwidth]{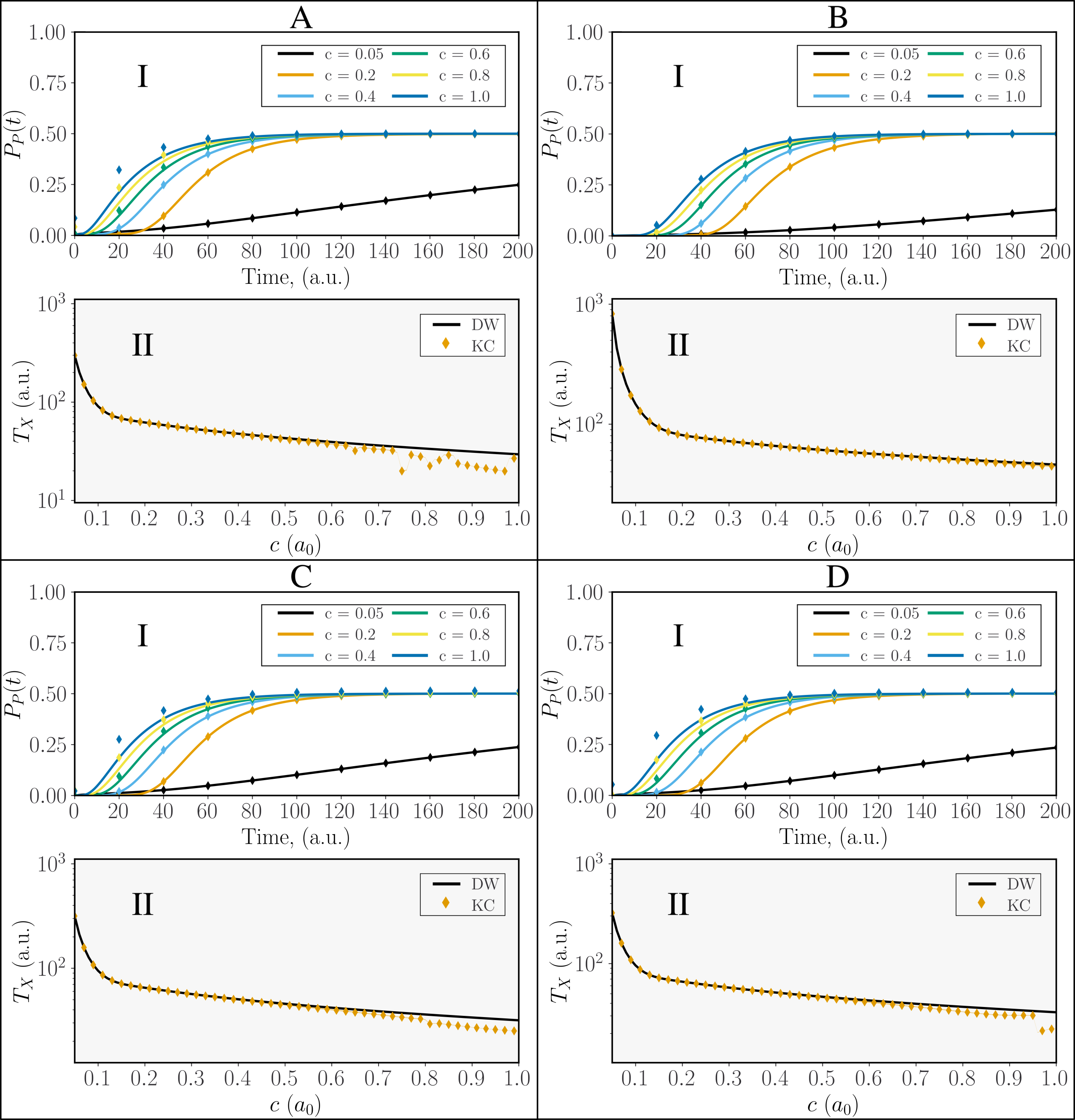}
    }
    \caption{Comparison of observables computed obtained with $\hat{H}_{DW}$ (solid lines) and $\hat{H}_{KC}$ (diamonds) as a function of $c$, using $\kappa=0.1$, and $\nth=0.1$. The time evolution of the product population for the four proton transfer reactions are shown in the top subpanels. The corresponding inverse reaction rate constants are shown in the bottom subpanels. These results were generated using 100 Fock basis for Hamiltonian and initial state preparation and a subspace of 50 eigenfunctions for dynamics propagation.
    }
    \label{fig:lindbladian_dynamics_k0.1_n0.1_NbF_100}
\end{figure}
\clearpage
\begin{figure}[ht!]
    \centering
    \makebox[\textwidth][c]{
        \includegraphics[width=0.95\textwidth]{
        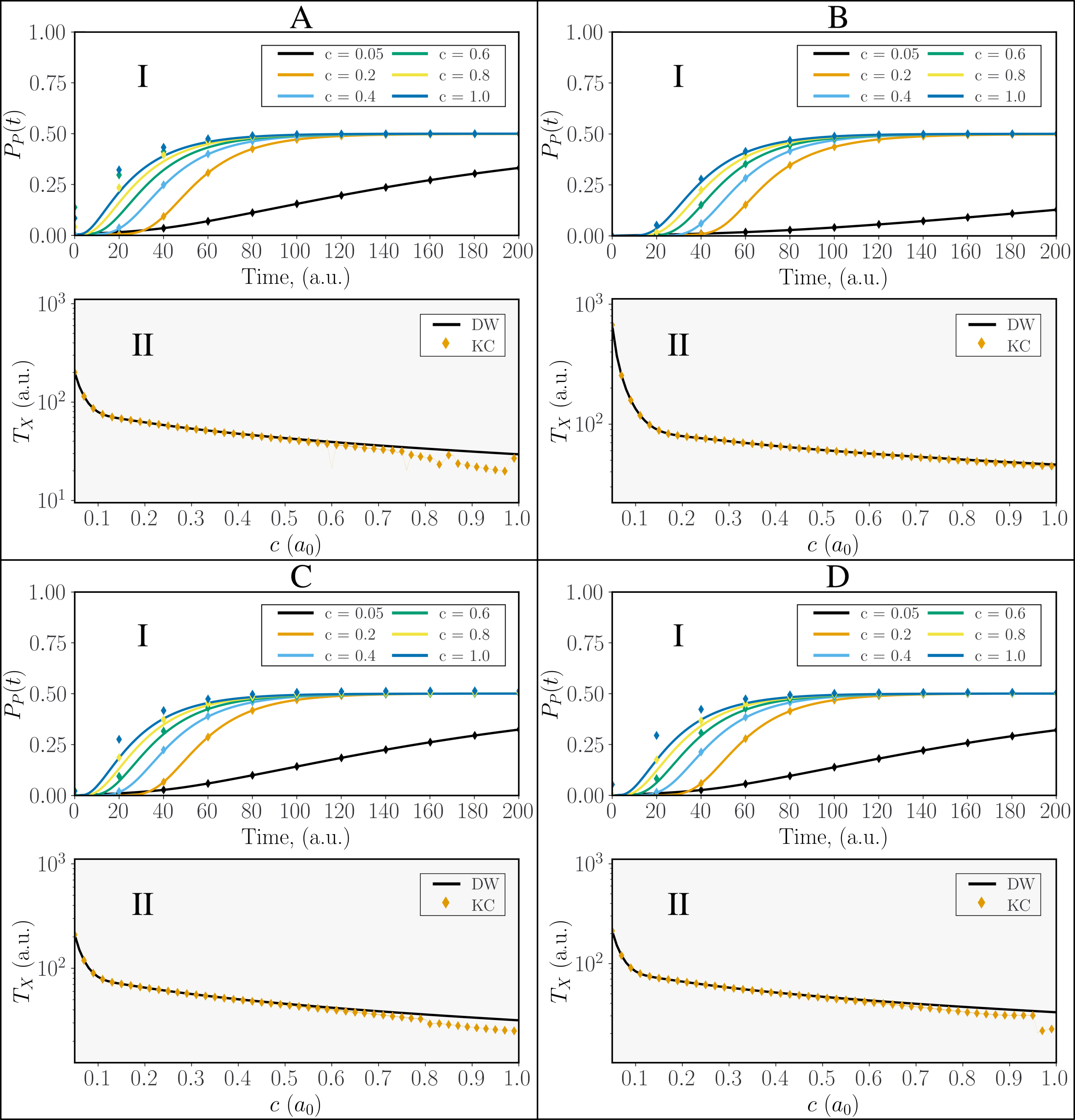}
    }
    \caption{Comparison of observables computed obtained with $\hat{H}_{DW}$ (solid lines) and $\hat{H}_{KC}$ (diamonds) as a function of $c$, using $\kappa=0.1$, and $\nth=0.1$. The time evolution of the product population for the four proton transfer reactions are shown in the top subpanels. The corresponding inverse reaction rate constants are shown in the bottom subpanels. These results were generated using 150 Fock basis for Hamiltonian and initial state preparation and a subspace of 50 eigenfunctions for dynamics propagation.
    }
    \label{fig:lindbladian_dynamics_k0.1_n0.1_NbF_150}
\end{figure}
\clearpage